\documentclass[11pt]{article}
\usepackage{amssymb,amsthm,amsmath,float}

\usepackage[total={6.5in,8.75in}, top=1.2in, left=0.9in, includefoot]{geometry}
\usepackage{graphicx}
\usepackage[pdftex,bookmarks, colorlinks, citecolor =blue, breaklinks]{hyperref}
\usepackage{url}

\newcommand{\Eq}[1]{(\ref{eq:#1})}

\newcommand{\Sec}[1]{\S \ref{sec:#1}}
\newcommand{\Fig}[1]{Fig.~\ref{fig:#1}}

\newcommand{\App}[1]{Appendix~\ref{app:#1}}

\newcommand{\Hyp}[1]{Hyp.~\ref{hyp:#1}}
\newcommand{\Hyps}[2]{Hyps.~\ref{hyp:#1}-\ref{hyp:#2}}

\newcommand{\InsertFig}[4]
{\begin{figure}[h!t]
       \centerline{
         \includegraphics[width=#4]{./figures/#1}
       }
       \caption{{\footnotesize  #2}
       \label{fig:#3}}
\end{figure}}

\newcommand{\InsertFigTwo}[5] {
\begin{figure}[h!t]
       \centerline{
         \includegraphics[width=#5]{./figures/#1}
         \hskip 0.5in
         \includegraphics[width=#5]{./figures/#2}
       }
       \caption{{\footnotesize  #3}
       \label{fig:#4}}
\end{figure}}

\newcommand{\InsertFigTwoVert}[5] {
\begin{figure}[h!t]
    \centerline{
	\renewcommand{\arraystretch}{0.01}
    \begin{tabular}{c}
     	\includegraphics[width=#5]{./figures/#1}\\
         \includegraphics[width=#5]{./figures/#2}
         \end{tabular}
       }
       \caption{{\footnotesize  #3}
       \label{fig:#4}}
\end{figure}}

\newcommand{\bC}{{\mathbb{ C}}}

\newcommand{\bR}{{\mathbb{ R}}}
\newcommand{\bS}{{\mathbb{ S}}}

\newcommand{\cO}{{\cal O}}

\newcommand{\eps}{\varepsilon}

\newcommand{\sgn}{\mathop{\rm sgn}\nolimits}

\newtheorem{hyp}{Hypothesis}

\newcommand{\beq}[1]{\begin{equation}\label{eq:#1}}
\newcommand{\eeq}{\end{equation}}

\newenvironment{se}[1]{\equation\label{eq:#1}\aligned}{\endaligned\endequation}
\newcommand{\bsplit}[1]{\begin{se}{#1}}
\newcommand{\esplit}{\end{se}}

\newenvironment{example}[1][]
  {
	\setlength \leftmargini {1.0em}		
	\setlength \topsep {0.5em}			
	\begin{quote}
	{\it Example#1} }
	{\end{quote}
  }
\newcommand{\bexam}[1][:]{\begin{example}[#1]}
\newcommand{\eexam}{\end{example}}

\title{Hamiltonian Triplet Interactions: Areal and Perimetric Forces}
\author{J.D.~Meiss\thanks
      {
       Useful conversations with Nathan Duignan,  Robert Easton, Richard Montgomery, 
       and Juan Restrepo are gratefully acknowledged. 
      }
    \\
 \begin{tabular}{cc}
	Department of Applied Mathematics\\
    University of Colorado \\
	Boulder, CO 80309-0526 \\
	James.Meiss@colorado.edu\\ 
\end{tabular}
}
\date{\today}
\begin{document}
\maketitle

\begin{abstract}
\vspace*{1ex}
\noindent

Gravitational and electromagnetic interactions are Hamiltonian systems with forces between pairs of particles.
We propose an alternative: Hamiltonian dynamics with triplet interactions between point particles. 
Our system has a potential energy that depends on the shape of the triangle for each triplet.
Similar multi-body forces occur in many physical systems, e.g., polarizable molecules, nucleon interactions, and colloids,
but typically are combined with more conventional two-body forces.
We focus on potentials that depend only on the triangle perimeter or on its area. The resulting  
forces point towards a center of the triangle, either the incenter or the orthocenter, respectively.
For the planar case, the resulting system has six degrees of freedom but can be reduced to
three since it conserves the total momentum and angular momentum.
The dynamics often exhibits chaotic motion, but there are a number of special solutions, 
for example equilateral and isosceles triangles, and perturbations of these can lie on invariant tori.
Numerical investigations of several examples show families of such regular trajectories as well as
examples of chaotic dynamics.
\end{abstract}

\section{Introduction: Triplet Interactions}\label{sec:Introduction}
A classic dynamical system corresponds to a set of particles with positions and momenta $(r_i,p_i) \in \bR^3 \times \bR^3$
and masses $m_i$, where $i = 1,2, \ldots n$. Its dynamics is  obtained from a Hamiltonian 
$H(r,p): \bR^{3n} \times \bR^{3n} \to \bR$ that has the standard  kinetic plus potential form:
\beq{StdHamiltonian}
	H(r,p) = K(p) + V(r) = \sum_{i=1}^n \frac{\|p_i\|^2}{2 m_i} + V(r) .
\eeq
For standard models, such as gravitational or electrostatic interactions, the particles interact pairwise with a central force so that 
\beq{twoBody}
	V(r) 
	     = \sum_{j<i} U_2(\|r_{ij}\|)
\eeq
where
\[
	r_{ij} \equiv r_i - r_j,
\]
The pair potential, $U_2$, can also depend on parameters, such as charge and mass, associated with each particle.
Instead of a potential of the form \Eq{twoBody}, we will consider cases where the interactions correspond to a force among triplets.
We assume the potential satisfies three hypotheses:
\begin{hyp}[Central Forces]\label{hyp:central}
The potential energy, $V(r)$, depends only on pairwise distances, $\|r_{ij}\|$, $i,j =1 \ldots n$.
\end{hyp}

\noindent
Of course this is the case for gravitational and electrostatic interactions, \Eq{twoBody}.
However, unlike this case, we will assume:
\begin{hyp}[Triplet Interactions]\label{hyp:triplet}
The potential is a sum of interactions between triplets:
\[
	V(r)  = \sum_{i<j<k} U_3(r_i,r_j,r_k).
\]
Moreover, $U_3: \bR^{9} \to \bR$ \emph{cannot} be written as the sum of pairwise interactions.
\end{hyp}

\noindent
Potentials that satisfy \Hyps{central}{triplet}, depend only on the lengths of the sides of the triangle 
formed by each triplet of particles at positions $(r_i,r_j,r_k)$ in $\bR^3$, thus
\beq{CentralForce}
	V(r)  = \sum_{i<j<k} U_\Delta(\|r_{ij}\|,\|r_{jk}\|,\|r_{ki}\|).
\eeq
For example if there are three bodies with positions $(r_1,r_2,r_3) = (u,v,w)$, and
\beq{Sides}
	a = \|w-v\|, \quad b = \|u - w\|, \mbox{ and } c = \|v-u\|, 
\eeq
(see e.g., \Fig{TriangleNotation}) then  \Eq{CentralForce} becomes

\beq{UDelta}
	V(u,v,w) = U_\Delta(a,b,c) ,
\eeq
so that $U_\Delta$ is a function only of the lengths of the sides of the $(u,v,w)$-triangle.
Of course, $U_\Delta$ could also depend on parameters, such as masses or charges.

Perhaps the simplest potential satisfying \Hyp{central} is that for harmonic springs:
$
	U_\Delta(a,b,c) = \tfrac12 \left( k_1 a^2 + k_2 b^2 + k_3 c^2 \right) .
$
However, this potential as well as those for gravitational or electrostatic forces
do not satisfy \Hyp{triplet}, since they are sums of pairwise interactions.

To make the triangular nature of the interactions more explicit, we assume:
\begin{hyp}[Symmetry]\label{hyp:symmetry}
The potential $U_\Delta: \bR^{3+} \to \bR$ is permutation symmetric under particle exchange:
\beq{Symmetric}
	U_\Delta(a.b.c) = U_\Delta(b,c,a) =U_\Delta(b,a,c).
\eeq
\end{hyp}

\InsertFig{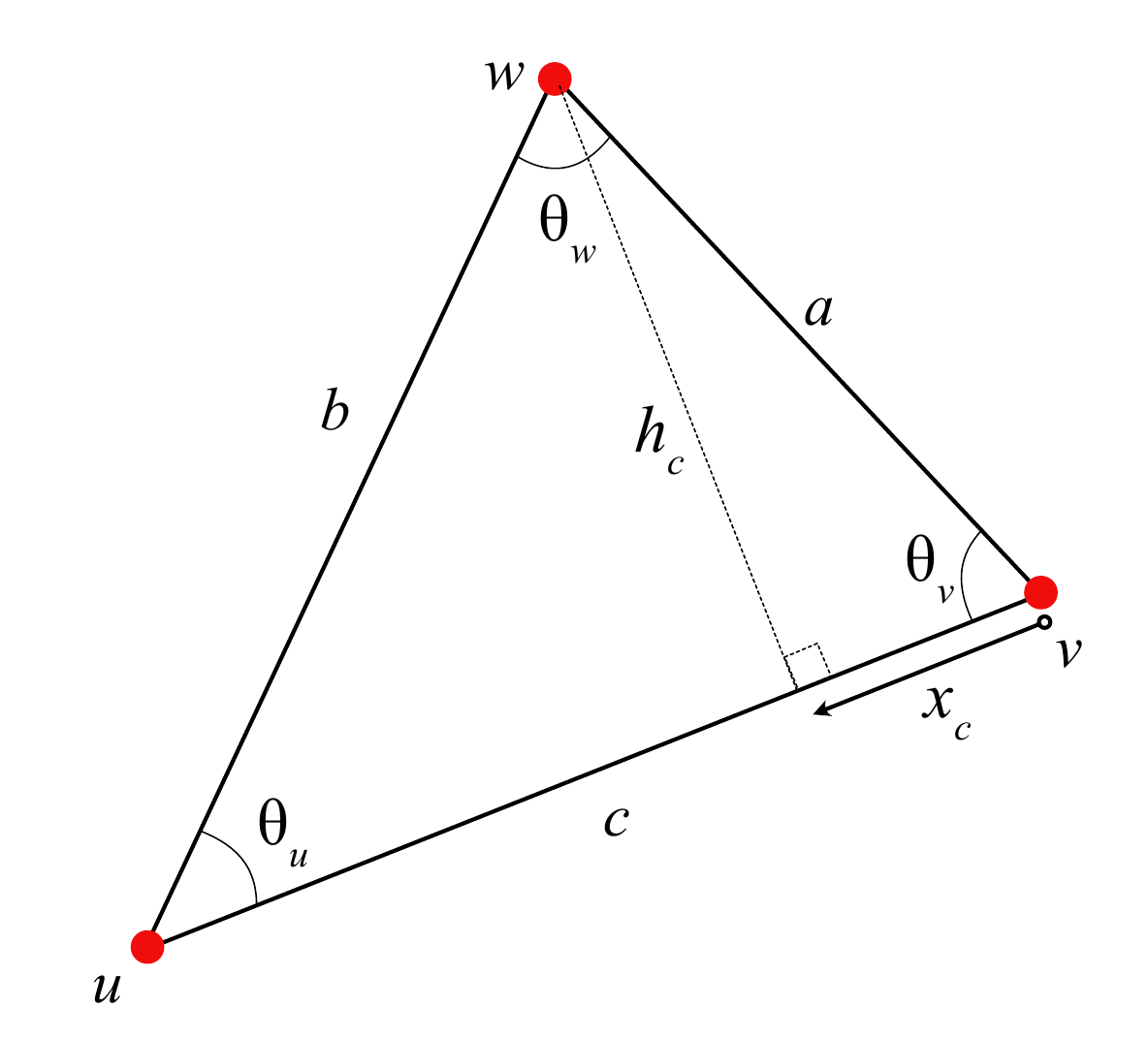}
{A triangle with vertices $u,v,w$, sides $a,b,c$, and interior angles $\theta_u,\theta_v,\theta_w$.
The altitude from side $c$ has length $h_c$ \Eq{Height},
and $x_c$ \Eq{xcsolve} denotes the distance from $v$ to the intersection
of this altitude with side $c$.}
{TriangleNotation}{3in}

For example, note that if the masses were all equal, a gravitational potential would satisfy \Hyp{symmetry};
however, it would not satisfy \Hyp{triplet}.

Potentials satisfying \Hyps{central}{symmetry} have been studied in the context of a number of applications, including
polarizable molecules, colloids, Bose-Einstein condensates, and nucleon interactions, as we recall in \Sec{Applications}.

In this paper we will study in detail only the simplest case: a single triplet
with positions $(u,v,w)$ and conjugate momenta $(p_u,p_v,p_w)$.
The Hamiltonian \Eq{StdHamiltonian} with a potential satisfying \Hyps{central}{symmetry} then becomes
\beq{Hamiltonian}
	H(r,p) = \frac{\|p_u \|^2}{2m_u} +\frac{\|p_v \|^2}{2m_v}+\frac{\|p_w \|^2}{2m_w}
			+ \ U_\Delta(\|v-w\|, \|w-u\|,\|u-v\|) .
\eeq
This gives the dynamics
\bsplit{EqTBP}
			\dot{p}_u &= - \partial_u U_\Delta 
			          =  -  \frac{u-w}{\|u-w\|} \partial_2 U_\Delta -  \frac{u-v}{\|u-v\|} \partial_3 U_\Delta\\
			\dot{u}   &= \frac{p_u}{m_u}		
\esplit
with cyclic permutations for the other particles. We will rewrite these equations for specific
potentials in \Sec{ThreeBodies}, and in Jacobi coordinates in \Sec{Jacobi}.

The Hamiltonian system \Eq{Hamiltonian} with three particles in $\bR^3$ has nine degrees of freedom. 
Of course since the potential \Eq{UDelta} has translation symmetry,
$V(u+\alpha,v+\alpha,w+\alpha) = V(u,v,w)$, $\forall \alpha \in \bR^3$, the total momentum
\beq{totalMomentum}
	p_T = p_u + p_v + p_w
\eeq
is conserved, and this can be used to eliminate three degrees of freedom by going to center-of-mass coordinates,
i.e., translating variables by 
\beq{CM}
	R = \frac{1}{m_T} (m_1 u + m_2 v + m_3 w)
\eeq
In addition, since \Eq{UDelta} has rotation symmetry: $V(Qu,Qv,Qw) = V(u,v,w)$ for any rotation matrix $Q \in SO(3)$,
the total angular momentum
\beq{JTotal}
	J_T = u \times p_u + v \times p_v + w \times p_w
\eeq
is also conserved.
It is also useful to note that moment of inertia relative to the center of mass,
\beq{InertiaForm}
	I =  \sum_{i=1}^3 m_i \|r_i-R\|^2 
	 = \frac{1}{m_T} (m_2 m_3 a^2 + m_1 m_3 b^2 + m_1 m_2 c^2) ,
\eeq
is also a function only of the triangle sides.

Note that the forces in \Eq{EqTBP} lie in the plane of the triangle formed by $(u,v,w)$. Thus if the initial momenta are in this plane, the orbits remain in the plane. For the planar case, we start with $3 \times 2$ degrees of freedom, and eliminate two by going to center-of-mass coordinates and one more for the angular momentum, as it is orthogonal to this plane. The resulting dynamical system has three degrees of freedom, and this can be made explicit using Jacobi coordinates, see \Sec{Jacobi}. 

Thus the dynamics of this simplest three-body case, unlike the simplest two-body case---the Kepler problem, is non-trivial.
We will show results of numerical simulations in \Sec{Simulations}.

\section{Multi-Body Forces in Applications}\label{sec:Applications}

Multi-body interactions have been considered for a number of different physical systems. One example is
for interactions between polarizable molecules \cite{Stone13}. 
A model of this is the Axilrod-Teller-Muto triple-dipole interaction \cite{Axilrod43}
(also obtained by Muto in 1943) with the potential
\beq{ATMPotential}
	 U_{ATM} = Z \sum_{i<j<k} \frac{1+3 \cos(\theta_i) \cos(\theta_j) \cos(\theta_k)}
	 				{(\|r_{ij}\|\|r_{ik}\|\|r_{jk}\|)^3},
\eeq
where, for example, $\theta_i$ is the angle between the sides $r_{ij}$ and $r_{ik}$ of the $ijk$ triangle.
Note that the potential \Eq{ATMPotential} is positive for equilateral and right triangles, but is negative when the particles lie on a line, so that one angle is $\pi$.

The ATM potential appears to not be of the form \Eq{CentralForce}, since it depends
on the angles. However, since the angles are determined by the sides, 
as we recall explicitly in \Eq{Cosines} in \App{Triangles}, \Eq{ATMPotential} satisfies \Hyps{central}{symmetry}.

Examples of interactions of micro-clusters were studied in \cite{Halicioglu80, Oksuz82} using 
a potential of the form $V = U_{LJ} + U_{ATM}$, with the Lennard-Jones two-body potential
\beq{LennardJones}
	U_{LJ} = \sum_{i<j} \left[ \frac{1}{\|r_{ij}\|^{12}}-\frac{1}{\|r_{ij}\|^{6}} \right] ,
\eeq
representing the Van der Waals forces.
Combining \Eq{LennardJones} and \Eq{ATMPotential} gives the ``LJAT'' interaction.
Minimum energy configurations for $3-6$ bodies as a function of the triplet strength $Z$
were studied in \cite{Halicioglu80};
for example when $Z$ is small the equilateral triangle
has minimum energy, but above a critical value, the linear configuration has smaller energy.
The paper \cite{Oksuz82} studies the stability of the icosahedral equilibrium, which is stable at $Z = 0$. 
The dynamics of this model is shown to be chaotic for a single ``trimer'' in \cite{Chakravarty97, Yurtsever97}.
The LJAT model is also discussed in the book by Stone \cite{Stone13}.

Another example corresponds to the interactions of polar molecules in an optical lattice \cite{Buchler07}; in this
case there are two-body interactions together with a three-body potential for which one model is
the Hubbard potential
\[
	U_H  =  \sum_{i<j<k} \left(\frac{1}{\|r_{ij}\|^3 \|r_{jk}\|^3} + \mbox{cyclic permutations} \right) .
\]

An alternative form for multi-body interactions was used by
Baskes \cite{Baskes99} to study crystalline lattices:
\[
	V  =  U_{LJ} +  \sum_{i=1} \bar{\rho}_i( \ln(\bar{\rho}_i) - 1) ,
\]
where $\bar{\rho}_i$ is a background electron density for the $i^{th}$ particle due to its $Z$ neighbors defined by
\[
	\bar{\rho}_i = \frac{1}{Z}\sum_{j \neq i} \exp(-\beta (\|r_{ij}\| -1)) .
\]
This approach is used to study lattices, and is a version of the ``Embedded Atom Method'' \cite{Daw93}.
For the three body case, where $Z = 2$, the multi-body portion of this potential satisfies \Hyps{central}{symmetry}.
A spine fit to three-body forces was used to obtain similar potentials for a Vanadium lattice in \cite{Lipnitskii16}.
 
Multi-body forces are also relevant in colloids, due to the nonlinearities in the Poisson-Boltzmann equations. 
In \cite{Russ02} the authors consider  hard spheres in a salt solution and numerically compute---to lowest order in an expansion---the effects of the screening due to the fluid on the interaction between the spheres. Empirically they find that the three-body interactions closely fits a potential of the Yukawa form
\beq{Yukawa}
	U_Y = - \sum_{i<j<k} \frac{e^{-\gamma ( \|r_{ij}\| + \|r_{jk}\| + \|r_{ki}\|)}}{ \|r_{ij}\| + \|r_{jk}\| + \|r_{ki}\|} ,
\eeq
so that it depends only upon the perimeter of each triangle. For this system, there is also a two-body, repulsive, Yukawa potential.  In \Sec{Perimeter}, we consider similar potentials that depend only upon the perimeter.

Other examples of multi-body forces include the quantum physics of BECs \cite{Kohler02,Johnson09}. 
where the potential depends only upon inter-particle distances.
Three-nucleon forces are also important for the nuclear stability \cite{Fukui24}, describe interactions with mesons
and pions \cite{Smith60, Hammer15,Carlson83} and arise in ``chiral effective field theory'' \cite{Hebeler21}.

In the context of non-Hamiltonian and dissipative systems, multi-particle forces have been used
in generalizations of the Kuramoto model
often used to study synchronization in networks of coupled oscillators, see e.g., \cite{Strogatz00}.
Multi-particle interactions correspond, in this case, to interactions on hyper-graphs or simplicial complexes \cite{Landry20,Zhang23}. For example, \cite{Lohe22} considered multi-body interactions between higher-dimensional phase oscillators. This non-Hamiltonian system has oscillators with positions $r_i \in \bS^2 = \{r \in \bR^3: \|r\| = 1\}$ and frequencies $\omega_i$.
For triplet interactions, Lohr uses the potential
\[
	U_L = \sum_{i,j,k} \eps_{ijk} r_i\cdot r_j \times r_k ,
\]
where $\eps_{ijk}$ is completely antisymmetric.
Upon adding the constraint, using Lagrange multipliers, that the particles remain on the sphere, the
dissipative dynamics become
\[
	\dot{r_i} = \omega_i \times r_i  + 
		\frac{\lambda}{N^2} \sum_{j,k=1}^N \eps_{ijk} [ r_j \times r_k - r_i( r_i \cdot r_j \times r_k)]
\]
Similar models are also studied in \cite{Dai21}, but with a symmetric coupling, giving equations of the form
\[
	\dot{r_i} = \omega_i \times r_i + \frac{\lambda}{N^2} \sum_{j,k=1}^N r_j \cdot r_i ( r_k -(r_k \cdot r_i) r_i) ,
\]
as well as higher dimensional cases. These interactions do not satisfy \Hyp{central}.

In \cite{Kim22} the gradient flow with a potential that depends upon the volume of a simplex is considered.
For example, for triplet interactions they use
\[
	U_\Delta = \frac{\kappa_2}{6N^2} \sum_{i,j,k=1}^N \|A(r_i,r_j,r_k)\|^2 ,
\]
where $A(u,v,w)$ is the (vector) area of the triangle with vertices $u,w,w$ (see \Eq{Area}). 
Using Heron's formula \Eq{Heron} for the area (see \App{Triangles}), the resulting equations become
\[
	\dot{r}_i = \frac{\kappa_2}{8N^2} \sum_{j,k=1}^N \left(\|r_{jk}\|^2(r_j+r_k-2r_i) + 
	(\|r_{ik}\|^2-\|r_{ij}\|^2)(r_j-r_k)\right) .
\]
Since this is a gradient system, the dynamics implies that areas collapse to zero, so the particles will collapse to a line.  We will assume the potential is a function of the area  in the Hamiltonian context in \Sec{Areal}.

\section{Three Body Forces}\label{sec:ThreeBodies}

In the remainder of this paper we will treat two special cases of the three-body potential \Eq{UDelta} that satisfies \Hyps{central}{symmetry}. For simplicity, we assume that there are only three-body forces,
neglecting the two-body terms that appear in most of the applications in \Sec{Applications}.
Denote the positions of the three particles by $u,v,w \in \bR^3$ and their momenta by
$p_u,p_v,p_w \in \bR^3$, the Hamiltonian is \Eq{Hamiltonian}

In \Sec{Perimeter}, we assume that the potential, $U_\Delta$, is a function only of the triangle's perimeter,
and in \Sec{Areal}, only of its area.

\subsection{Perimetric Forces }\label{sec:Perimeter}

Suppose first that $U_\Delta = U(P)$, depends only on the \textit{perimeter} of the triangle
(recall \Eq{Yukawa}),
\beq{Perimeter}
	P(u,v,w) = \|w-v\| + \|u-w\| + \|v-u\|  = a+b+c \;,
\eeq
where $a,b,c$ are the side lengths seen in \Fig{TriangleNotation}.
In the simplest case $U(P) = P$; however, this does not satisfy \Hyp{triplet} since it is a sum of pair interactions.
We considered several cases, but concentrate on the simplest, $U(P) = \tfrac12 P^2$.

The resulting equations of motion \Eq{EqTBP} become
\bsplit{PerimEq}
	\dot{p}_u & 
		        =  U'(P)\left( \frac{v-u}{c} + \frac{w-u}{b}\right) ,\\
	\dot{u} &= \frac{p_u}{m_u} ,
\esplit
with cyclic permutations for the other two particles.

Note that the force vectors in \Eq{PerimEq} bisect the interior angles of the triangle, 
pointing into its interior if $U'(P) > 0$, see \Fig{PerimeterForces}.
These three force vectors define lines that meet at the so-called \textit{incenter} of the triangle,
\beq{incenter}
 c_i = \frac{au + bv+ cw}{P} ,
\eeq
the center of the circle inscribed in the triangle, see \App{Incenter}. To see this, note that the two right triangles formed from a particle's force vector with each of the triangle edges coming from that particle are congruent: i.e., the three yellow edges in \Fig{PerimeterForces} have equal lengths and are radii of the inscribed circle.

If the particles become collinear at any time, then the middle particle is at the incenter. 
Indeed when the particles are collinear, then the perimeter is twice the maximum distance.
For example, if $v$ is between  $u$ and $w$, then $a+c = b$ and $P = 2b$.
In this case $c_i =  v$, the incenter \Eq{incenter} is the position of the central particle.

\InsertFig{PerimeterForces}{Sketch of the forces acting on a triple of particles at $u,v,w$ when the potential is a function of the
perimeter of the triangle. By \Eq{PerimEq} force vectors bisect the angles at each vertex, and thus meet at the incenter of the triangle when $U'(P) > 0$, see \App{Incenter}.}
{PerimeterForces}{3in}

The rate of change of the perimeter of the triangle is
\[
	\frac{d}{dt} P=   \frac{1}{a} (v-w) \cdot (\dot{v}-\dot{w})  
	                + \frac{1}{b} (w-u) \cdot (\dot{w}-\dot{u}) 
	                + \frac{1}{c} (u-v) \cdot (\dot{u}-\dot{v})
\]
Combining this with \Eq{PerimEq} gives
\[
	\frac{d}{dt} U(P) = {U'(P)} \frac{d}{dt} P 
					   = -\left(\dot{u} \cdot \dot{p}_u + \dot{v}\cdot \dot{p_v} + \dot{w} \cdot \dot{p_w} \right)
				       = - \frac{d}{dt} K
\]
where $K$ is the kinetic energy, e.g. \Eq{Hamiltonian}: energy is conserved.

Using \Eq{incenter},  rate of change of the incenter is

\begin{align*}
	\frac{d}{dt} c_i &= \frac{d}{dt} \frac{au + bv+ cw}{P} \\
	               &=  \frac{1}{P} \left(- c_i \dot{P} +  a\dot{u} + b \dot{v} + c \dot{w} + 
	                       \frac{u}{a}(v-w)\cdot (\dot{v}-\dot{w}) +
	                       \frac{v}{b}(w-u)\cdot (\dot{w}- \dot{u}) +
	                       \frac{w}{c}(u-v)\cdot (\dot{u}-\dot{v}) \right) 
\end{align*}


\subsection{Areal Forces}\label{sec:Areal}

Suppose now that $U_\Delta$ depends only on the \textit{area} of the triangle formed by a triplet.
We denote the (vector) area of a triangle by
\bsplit{Area}
	A(u,v,w) &=  \tfrac12 (v-u) \times (w-u) \\  
			 &= \tfrac12 (u \times v + v \times w + w \times u) .
\esplit
This vector is normal to the face of the triangle;  in 2D 
$A = \hat{e}_3  \|A(u,v,w) \|$ when $u,v,w$ are a right-handed triplet.
The form \Eq{Area} is invariant under even permutations of the vertices, and its magnitude is permutation invariant:
\[
	 \|A(u,v,w) \| =  \|A(v,u,w) \| =  \|A(w,v,u) \|.
\]
Moreover, the area is a function only of the lengths of the sides of the triangle, as can be seen most
explicitly in Heron's formula \Eq{Heron} in \App{Triangles}.
Thus a potential that depends only upon $ \|A \|$ obeys \Hyps{central}{symmetry}. 

The resulting equations of motion for the Hamiltonian \Eq{Hamiltonian} become
\begin{align*}
	\dot{p}_u &= - U'( \|A\|) \frac{1}{2 \|A \|} \nabla_{u}  \|A\|^2 ,\\
	\dot{u}   &= \frac{p_u}{m_u} ,
\end{align*}
with cyclic permutations. Using the summation convention and \Eq{Area}, the derivative of the squared area with respect to a component of $u$ becomes
\begin{align*}
	\frac{\partial}{\partial u_\alpha}  \|A \|^2 
	 &=  2 A_i  \frac{\partial A_i}{\partial u_\alpha} 
	 = \tfrac12 \epsilon_{ijk} (w_j-v_j)(u_k - v_k) \epsilon_{i\beta \alpha}(w_\beta-v_\beta) \\
	 &= \tfrac12 \left[(w_\beta - v_\beta)(u_\alpha-v_\alpha)(w_\beta-v_\beta) - 
	 	         (w_\alpha-v_\alpha)(u_\beta- v_\beta)(w_\beta-v_\beta)\right] \\
	&= \tfrac12 \left[(u_\alpha-v_\alpha) \|w-v \|^2 - (w_\alpha - v_\alpha) (u-v)\cdot(w-v) \right] .
\end{align*}
Thus in vector notation 
\begin{align*}
	\nabla_u  \|A \|^2 	&= \tfrac12 [(u- v)  \|w-v \|^2 -(w-v)((u-v)\cdot(w-v))] \\
					& = \tfrac12 (w-v) \times[ (u-v) \times (w-v)] \\
					& = (v-w) \times A .
\end{align*}
This vector is orthogonal to both the opposite side vector $v-w$ and to $A$:
\[
	(v-w) \cdot \nabla_u  \|A \|^2 =  A \cdot \nabla_u  \|A \|^2 = 0 .
\]
Consequently, the equations of motion become
\bsplit{ArealEqs}
	\dot{p}_u &= m_1 \ddot{u} =  -\tfrac12 U'( \|A \|) (v-w) \times\hat{A} , \\
	\dot{p}_v &= m_2 \ddot{v} =  -\tfrac12 U'( \|A \|) (w-u) \times \hat{A} ,  \\
	\dot{p}_w &= m_3 \ddot{w} =  -\tfrac12 U'( \|A \|) (u-v) \times \hat{A} ,
\esplit
where $\hat{A} = A/ \|A \|$ is the unit vector in the direction of the area.

As seen in \Fig{ArealForces} the resulting force vectors in \Eq{ArealEqs} are in the plane of the 
triangle and are orthogonal to the opposite side of the triangle: they are parallel to
the triangle's altitudes. As discussed in \App{Orthocenter}, the
altitudes have a common intersection,  the \textit{orthocenter} of the triangle.
When the triangle is acute, and $U' > 0$, the forces point inward, and the
orthocenter is in the interior of the triangle. However, when the triangle has an 
obtuse angle, two of the force vectors point outward, and the orthocenter is exterior
to the triangle.

\InsertFig{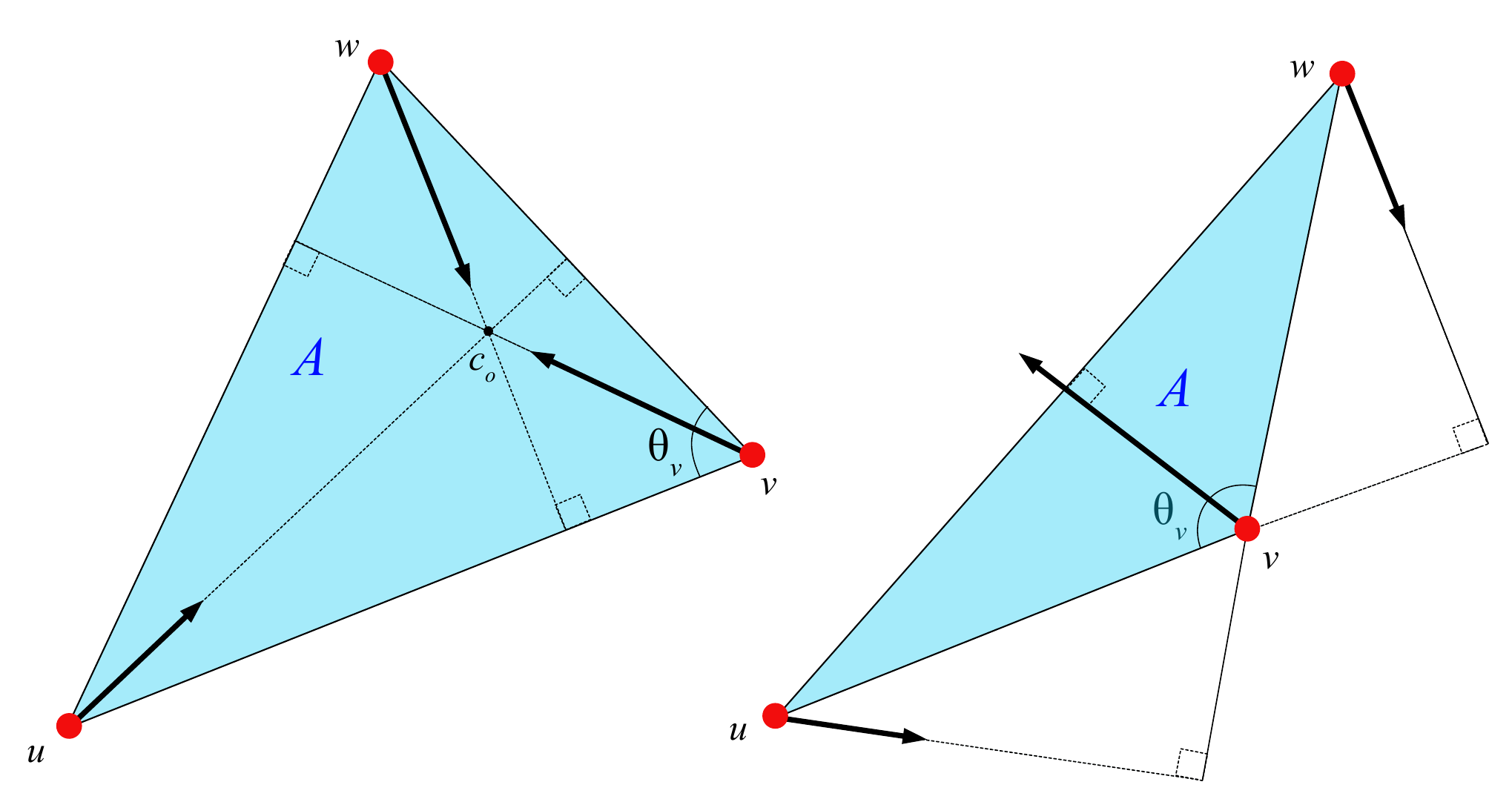}{Sketches of the forces acting on a triplet of particles at $u,v,w$ when the potential
is a function of the area. The forces are orthogonal to the opposite sides and meet at the orthocenter, $c_o$.
}{ArealForces}{5in}

Naturally, these equations also preserve the total momentum \Eq{totalMomentum} and total angular momentum \Eq{JTotal}.
Note that the rate of change of the area of the triangle is
\[
	\frac{d}{dt} A(u,v,w) = \frac12\left(\dot{u} \times(v-w) + \dot{v} \times(w-u) + \dot{w}\times(u-v)\right) .
\]
\subsection{Special Solutions}\label{sec:Special}

There are a number of special configurations for which the dynamics is simpler.
For simplicity we suppose that all the masses are equal
\beq{EqualMass}
	m_u = m_v = m_w = 1,
\eeq
the center of mass \Eq{CM} is the origin, and the total momentum \Eq{totalMomentum} is zero.

\begin{itemize}
\item (Equilateral Triangle) Suppose that the initial state has a discrete rotational symmetry:
\bsplit{EquilateralSym}
		v(0) &= Q_{2\pi/3} u(0), \quad w(0) = Q_{4\pi/3} u(0), \\
		\dot{v}(0) &= Q_{2\pi/3} \dot{u}(0), \quad\dot{w}(0) = Q_{4\pi/3} \dot{u}(0),
\esplit
for the 2D rotation by angle $\phi$, $Q_\phi$. This corresponds to an equilateral triangle with sides
\[
	a = b = c = \sqrt{3}\|u\|,  
\]
so that the perimeter and area become
\begin{align*}
	P &= 3^{3/2}\|u\| , \\
	A &= \tfrac{3^{3/2}}{4} \|u\|^2 .
\end{align*}
The rotational symmetry of the initial conditions is maintained by \Eq{PerimEq}, and so the shape remains equilateral.
Letting $u = r (\cos\theta, \sin\theta)$, the system can be reduced to a Hamiltonian with coordinates $(r,\theta)$, and  momenta $p_r = \dot{r}$ and $p_\theta = r^2 \dot{\theta}$:
\[
	H(r,\theta, p_r,p_\theta) = \tfrac12 p_r^2 + \frac{p_\theta^2}{2r^2} + \tfrac{1}{3} U ,
\]
Note that this is an integrable system, since the angular momentum is conserved.
Assuming that $U$ is monotonically increasing, the uniformly rotating case corresponds 
to $p_\theta^2 = \sqrt{3}U'(P)r^3$ or $\frac{1}{2\sqrt{3}}U'(A) r^4$, for the 
perimetric and areal cases, respectively.

\item (Isosceles) For an isosceles triangle with base $b$ and height $h$, we can set
$u(t) = (-b/2,-h/3)$, $v(t) = (0,2h/3)$ and $w(t) = (b/2,-h/3)$. 
This form is a solution of \Eq{PerimEq} maintaining the symmetry. Here
\begin{align*}
	P(t) &= |b|+  \sqrt{b^2+4h^2}, \\
	A(t) &= -\tfrac12 bh \hat{z}.
\end{align*}
Note that the area is negative for our assumed coordinates when $bh > 0$ since the orientation 
of the triangle is reversed.
The dynamics can be reduced to a two degree of freedom Hamiltonian
\[
	H(b,h,p_b,p_h) = p_b^2 + \tfrac34 p_h^2 + U(b,h) .
\]
For the perimetric case, the equations become
\begin{align*}
		\ddot{h} &= \tfrac32 \dot{p}_h = -6U'(P) \frac{h}{\sqrt{b^2 + 4h^2}} \\
		\ddot{b} &= 2 \dot{p}_b = -2 U'(P)\left(\sgn(b) + \frac{b}{\sqrt{b^2+4h^2}} \right) .
\end{align*}
Similarly the equations for the areal case are
\begin{align*}
	\ddot{b} &=  2\dot{p}_b = -U'(\|A\|) |h| \sgn(b)\\
	\ddot{h} &=  \tfrac32 \dot{p}_h = -\tfrac{3}{4}  U'(\|A\|) |b| \sgn(h) .
\end{align*}
These equations are not smooth for a pair collision when $b=0$, or
a linear configuration when $h=0$.

\item{(Rotating Line)} Suppose that the particles lie on a line, and one is at the center of mass,
say $v = 0$, so that $u = -w$. This linear configuration is preserved if 
$\dot{u}(0) = -\dot{w}(0)$ and $\dot{v} = 0$.
Using polar coordinates, $u = r (\cos(\theta), \sin(\theta))$, the perimeter is $P = 4r$, 
and the reduced Hamiltonian becomes
\[
	H(r,\theta,p_r,p_\theta) = \tfrac12 p_r^2 + \frac{p_\theta^2}{2r^2} + \tfrac12 U(4r) .
\]
Since the angular momentum is conserved, this again reduces to one degree-of-freedom.
If $U(P)$ is increasing, this has the uniformly rotating solution when
$
	p_\theta^2 =  2r^3 U'(P)  .
$
A special case of this corresponds to zero angular momentum. Again assuming $U(P)$ is monotone increasing, this gives oscillations with repeated triple collisions.
\end{itemize}

\section{Jacobi Coordinates}\label{sec:Jacobi}

To explicitly eliminate the conserved quantities we can use, for example, Jacobi coordinates \cite{Meyer92, Littlejohn97,Littlejohn98, Montgomery17}).
For the triplet $(r_1,r_2,r_3) = (u,v,w)$, one version of these coordinates is
\bsplit{Jacobi}
	s &= v-u \;, \\
	h &= w - \frac{m_1 u + m_2 v}{m_{12}} = \frac{m_1(w-u) + m_2(w-v)}{m_{12}} \;, \\
	R &=  \frac{1}{m_T} (m_1 u + m_2 v + m_3 w) \;, 
\esplit
where $m_T = m_1  + m_2  + m_3$ and $m_{12} = m_1 + m_2$. As sketched in \Fig{TriangularJacobi},  
the vector $s$ is one side of the triangle, $h$ is the vector from the center-of-mass of $u$ and $v$ to the vertex $w$,
and $R$ is the center-of-mass of the triplet. 
The inverse of \Eq{Jacobi} is
\bsplit{InverseJacobi}
	u &= R - \frac{m_2}{m_{12}}s - \frac{m_3}{m_T}h  \;,\\
	v &= R +\frac{m_1}{m_{12}}s - \frac{m_{3}}{m_T} h \;,\\
	w &= R + \frac{m_{12}}{m_T}h  \;.
\esplit

The main points of the choice \Eq{Jacobi} are:
(i) this linear transformation is orthogonal with respect to the mass-scaled metric,
so the kinetic energy remains diagonal, and
(ii) $(s,h)$ are invariant under translations, $(u,v,w) \to (u,v,w) + (\alpha,\alpha,\alpha)$ \cite{Montgomery17}.
Indeed the kinetic energy becomes
\beq{LKineticJacobi}
	K = \tfrac12 \left( \mu_1  \|\dot{s} \|^2 + \mu_2  \|\dot{h} \|^2 +  m_T  \|\dot{R} \|^2\right) \;,
\eeq
where
\[
	\frac{1}{\mu_1} \equiv \frac{1}{m_1} + \frac{1}{m_2} , \quad 
	\frac{1}{\mu_2} \equiv \frac{1}{m_{12}} + \frac{1}{m_3} \;,
\]
define the ``reduced masses''.
Since by \Hyp{central}, the potential is independent of the center-of-mass $R$, the momentum $p_R = m_T \dot{R}$ is conserved, and we can drop it from the Hamiltonian so that
\beq{HJacobi}
	H = \tfrac1{2 \mu_1}  \|p_s \|^2 + \tfrac1{2 \mu_2}  \|p_h \|^2 + U(s,h) \;.
\eeq
For the planar case, this is a four degree-of-freedom system.
The angular momentum \Eq{JTotal} becomes
\[
	J_T = s \times p_s +  h \times p_h ,
\]
which of course is conserved. Similarly the moment of inertia in center-of-mass coordinates 
\Eq{InertiaForm} becomes, in Jacobi coordinates,
\beq{IJacobi}
	I = \mu_1  \|s \|^2 + \mu_2 \|h \|^2 .
\eeq

\InsertFig{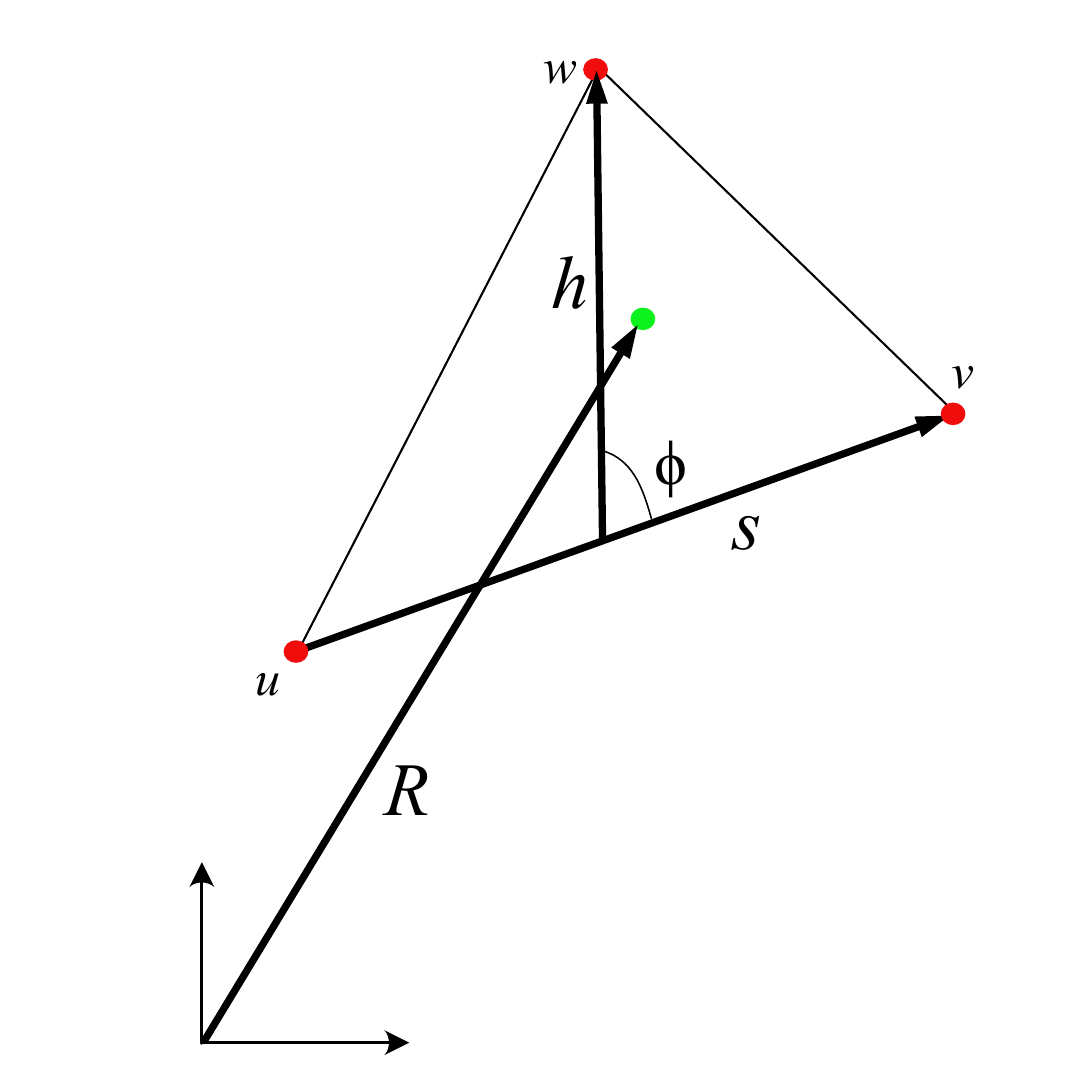}{Jacobi coordinates $(s,h,R)$ for the triangle.}{TriangularJacobi}{3in}

In Jacobi coordinates, the perimeter \Eq{Perimeter} becomes
\[
	P =   a + b + c =  \frac{1}{m_{12}} \left(\|m_1 s-m_{12}h \| +\|m_2 s +m_{12} h\| \right) + \|s\|.
\]
When the potential depends only on $P$,  the Hamiltonian equations for \Eq{HJacobi} are
\bsplit{JacobiP}
	\dot{p}_s & =  - U'(P)  \left( \frac{s}{\|s\|} + \frac{m_1}{m_{12}} \frac {m_1 s-m_{12}h}{\|m_1 s-m_{12}h\|}
								 +  \frac{m_2}{m_{12}} \frac{ m_2 s+m_{12} h}{\| m_2 s+m_{12} h\|} \right) ,\\
	\dot{p}_h & =  - U'(P) \left(  \frac {m_{12}h- m_1 s}{\|m_{12}h- m_1 s\|}
								 +  \frac{m_{12} h+ m_2 s}{\|m_{12} h+ m_2 s\|}\right) ,\\
	\dot s &= \tfrac{1}{\mu_1} p_s ,\\
	\dot h &= \tfrac{1}{\mu_2} p_h .
\esplit	

For the areal case, since $v-u = s$ and  $w-u = h + \tfrac{m_2}{m_{12}}s$, the area \Eq{Area} becomes
\[
	A(u,v,w) =  \tfrac12 s \times (h - \tfrac{m_2}{m_{12}}s) 	 
			 = \tfrac12 s \times h ,
\]
which is natural since $s$ is a side and $h$ is a vector connecting that side to the third vertex.
When the potential is a function of area, the equations for \Eq{HJacobi} are
\bsplit{JacobiA}
	\dot{p}_s & =  -\tfrac12 U'( \|A \|)  h \times \hat{A}  \\
	\dot{p}_h & =  -\tfrac12 U'( \|A \|)  \hat{A} \times s \\
	\dot s &= \tfrac{1}{\mu_1} p_s\\
	\dot h &= \tfrac{1}{\mu_2} p_h
\esplit


To eliminate the angular momentum we introduce polar coordinates following \cite{Montgomery17}. 
Using complex notation, $\bR^2 \simeq \bC$, let
\[
	s = \sigma e^{i\theta}, \quad h = \eta e^{i(\theta+\phi)} .
\]
Substituting into \Eq{LKineticJacobi} gives the kinetic energy
\[
	K = \tfrac12 \left(\mu_1 \dot{\sigma}^2 +  \mu_2 \dot{\eta}^2 +  
	\mu_1 \sigma^2 \dot{\theta}^2 + \mu_2\eta^2 (\dot{\theta} + \dot{\phi})^2 \right) .
\]
The Lagrangian $L = K -V$ does not depend upon the rotation angle $\theta$: 
the shape of the triangle is determined by $(\sigma,\eta, \phi)$.
This will give the conserved angular momentum,  $p_\theta = J_T$
\[
	p_\theta = \frac{\partial L}{\partial \dot{\theta}} 
	         = \mu_1 \sigma^2 \dot{\theta} +\mu_2\eta^2(\dot{\theta}+\dot{\phi})
\]
Since \Eq{IJacobi} gives $I = \mu_1 \sigma^2 + \mu_2 \eta^2$, eliminating $\dot{\theta}$ in the kinetic energy gives
\[
	K = \tfrac12 \left(\mu_1 \dot{\sigma}^2 +  \mu_2 \dot{\eta}^2 +  
	\frac{p_{\theta}^2}{I} +  
	\frac{\mu_1\mu_2\sigma^2\eta^2}{I}\dot{\phi}^2\right)
\]
Defining canonical momenta,
\begin{align*}
	p_\sigma = \mu_1 \dot{\sigma} , \quad 
	p_\eta = \mu_2 \dot{\eta} , \quad
	p_\phi =  \frac{\mu_1\mu_2\sigma^2\eta^2}{I}\dot{\phi} ,
\end{align*}
then gives
\beq{ReducedJacobi}
	H= \frac{p_\sigma^2}{2\mu_1} + \frac{p_\eta^2}{2\mu_2} + \frac{p_\theta^2}{2I} 
	+ \frac12 \left( \frac{1}{\mu_1\sigma^2} + \frac{1}{ \mu_2\eta^2} \right) p_\phi^2 
	+ U(\sigma,\eta,\phi)
\eeq
The potential is a function only of the lengths $\sigma$, $\eta$ and the angle $\phi$; therefore
this is a three degree-of-freedom system for $(\sigma, p_\sigma), (\eta, p_\eta), (\phi, p_\phi)$.

The perimeter and area become
\begin{align*}
	P &=  \frac{1}{m_{12}}(|m_{12}\eta e^{i\phi}-m_1\sigma | + |m_{12}\eta e^{i\phi}+m_2\sigma |) + \sigma \\
	A &= \tfrac12 \sigma\eta \sin(\phi)
\end{align*}

The configuration equations become
\bsplit{JacobiM}
	\dot{\sigma} &= \frac{1}{\mu_1} p_\sigma ,\\
	\dot{\eta}     &= \frac{1}{\mu_2} p_\eta ,\\
	\dot{\phi}     &= \left( \frac{1}{\mu_1\sigma^2} + \frac{1}{ \mu_2\eta^2} \right) p_\phi ,\\
\esplit
and for areal potentials the momentum equations are
\bsplit{JacobiRotating}
	\dot{p_\sigma} &= \mu_1\frac{p_\theta^2}{I^2}\sigma +\frac{p_\phi^2}{\mu_1 \sigma^3} - \tfrac12 U'( \|A \|) \eta |\sin(\phi)| ,\\
	\dot{p_\eta}   &=  \mu_2\frac{p_\theta^2}{I^2}\eta +\frac{p_\phi^2}{\mu_2 \eta^3} - \tfrac12 U'( \|A \|) \sigma |\sin(\phi)| ,\\
	\dot{p_\phi}   &= -\tfrac12 U'( \|A \|) \sigma \eta \cos(\phi) \,\mbox{sign}(\sin(\phi)) .
\esplit
The corresponding equations for the perimetric case are complicated, and we do not write them out.


\section{Numerical Explorations}\label{sec:Simulations}

A typical trajectory of the three body system \Eq{Hamiltonian} with $m_1=m_2=m_3 = 1$ and $U(P) = \tfrac12 P^2$ is shown
in \Fig{SidesPhase}. In panel (a), the particle positions are shown in the $(x,y)$ plane. The initial
conditions are indicated by filled circles, and curves for the three particles, red for $u$, green for $v$ and blue for $w$,
show the trajectories up to  $t=20$. The integration was done using the \emph{RK45} method in Matlab,
with an error bound of $10^{-12}$.
As one estimate of the integration accuracy, the energy and angular momentum have errors $\cO(5\times10^{-10})$
over an integration time of $t = 500$.
Panel (b) shows the same trajectory in the 3D space of inequities $(i_a,i_b,i_c)$ (see \App{Inequities}).
We claim these coordinates are more convenient than the three side lengths, $(a,b,c)$, since the triangle inequalities
restrict the latter to a hard-to-visualize cone in the positive octant, 
while the inequities can take any values in the non-negative octant. 

\InsertFigTwo{PerimeterSidesvpert0Phase}{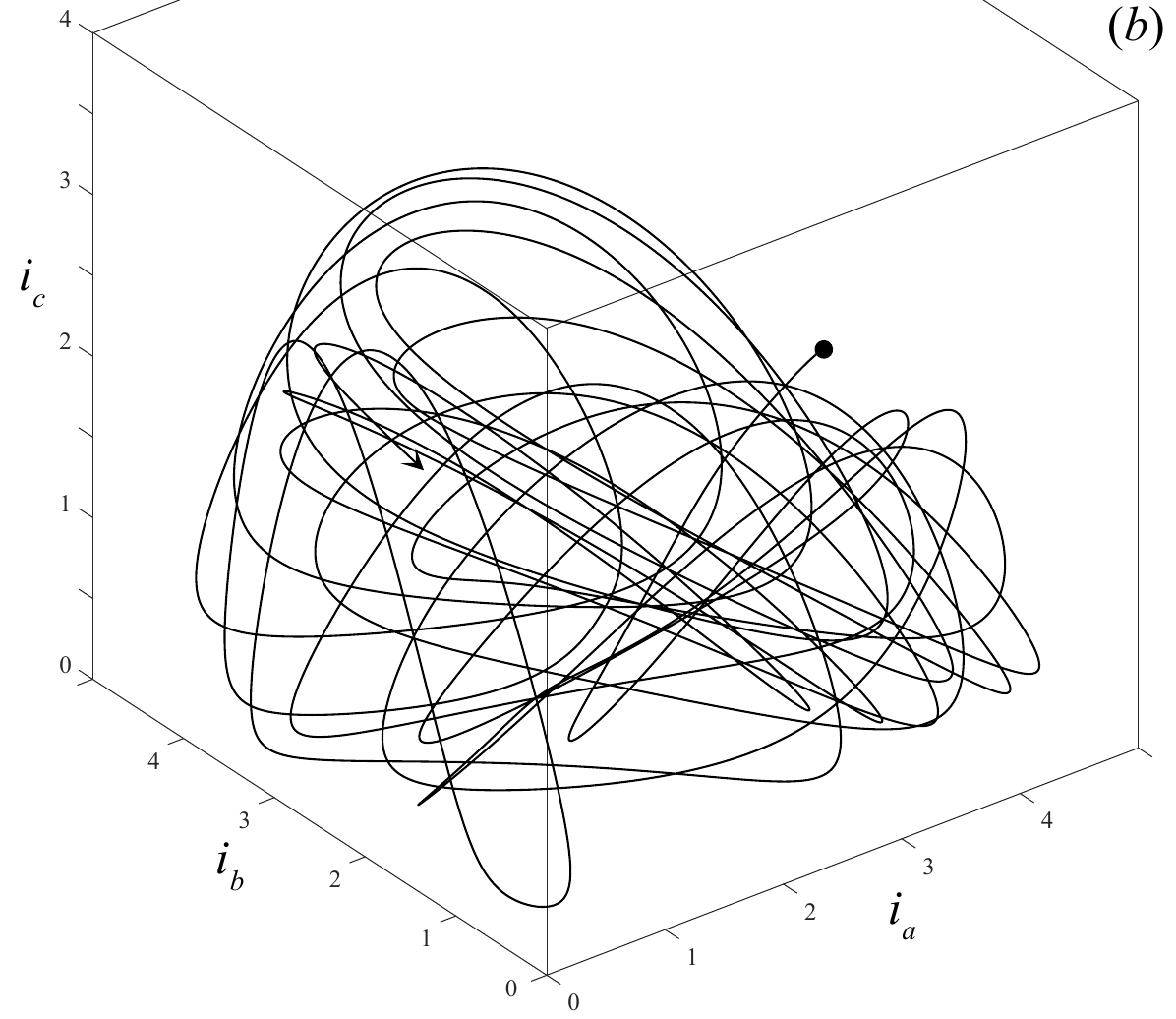}
{(a) Positions of the three particles $(u,v,w)$ in the plane as a function of time, 
up to $t = 20$ for $U(P) = \tfrac12 P^2$.
(b) Trajectory in the positive octant of the inequities, $(i_a,i_b,i_c)$, see \App{Inequities}.
At $t = 0$, the triangle has sides
$(a,b,c) = (2.9,2.0,1.8)$. For a center-of-mass at the origin this gives 
 $u_0 \simeq (-0.49166  -0.65781)$, 
 $v_0 \simeq (1.30833, -0.65781)$ and 
 $w_0 \simeq (-0.81666, 1.31561)$,
The initial velocities are $\dot{u} = -\dot{w} = (1,3)$ and  $\dot{v} = 0$.
For this trajectory, $E = 32.445$ and $J_T = 2.94842$.}
{SidesPhase}{3in}

If one plots longer segments of this trajectory, one gets the impression it is chaotic.
To see this in more detail, \Fig{SidesAreaPerim} shows $A(t)$ and $P(t)$ along this trajectory.
These curves are computed using \textit{RelTol} = \textit{AbsTol} $= 10^{-12}$; if this is relaxed,
the resulting curves for $A$ and $P$ appear close to those seen in the figure up to $t \approx 200$,
but after that the evolution is quite different, again as one would expect for a chaotic orbit.

\InsertFigTwoVert{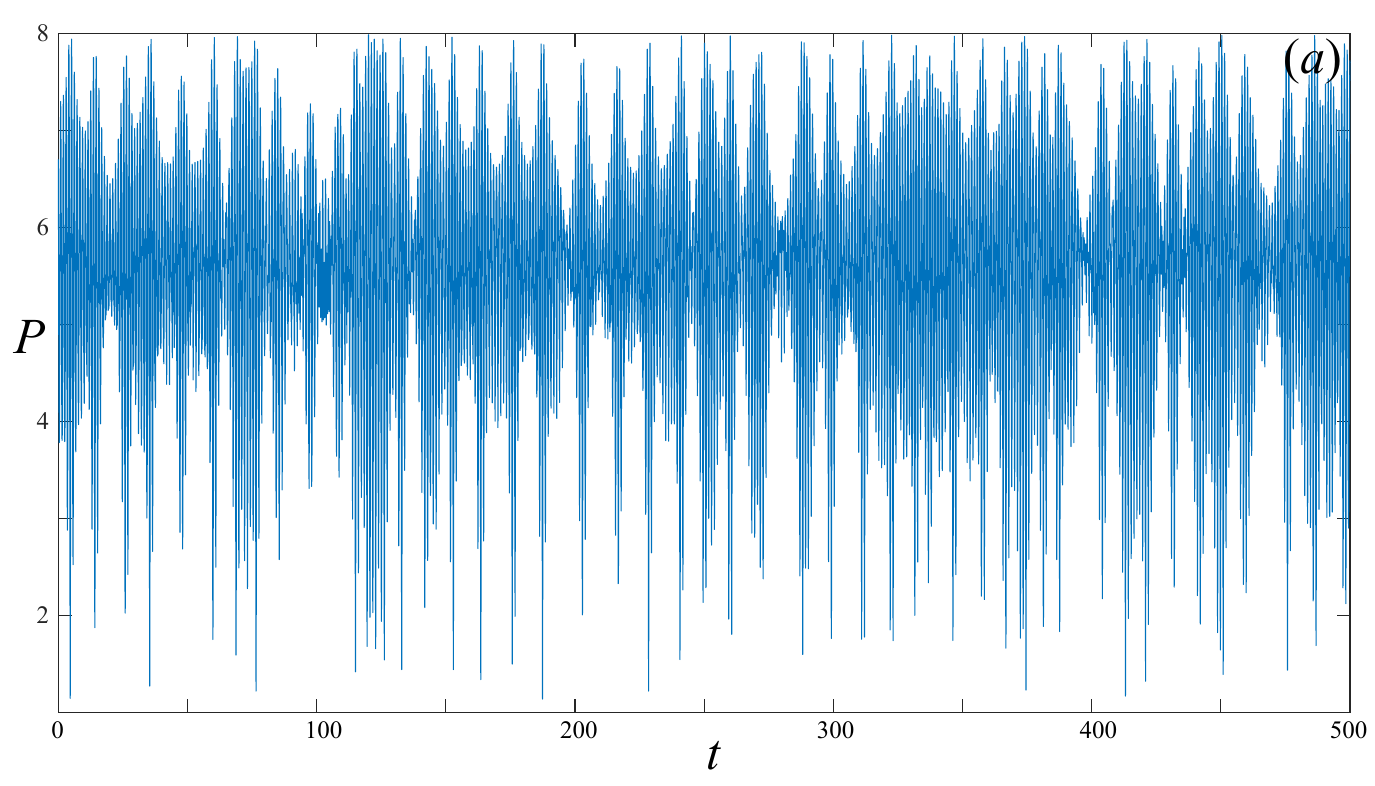}{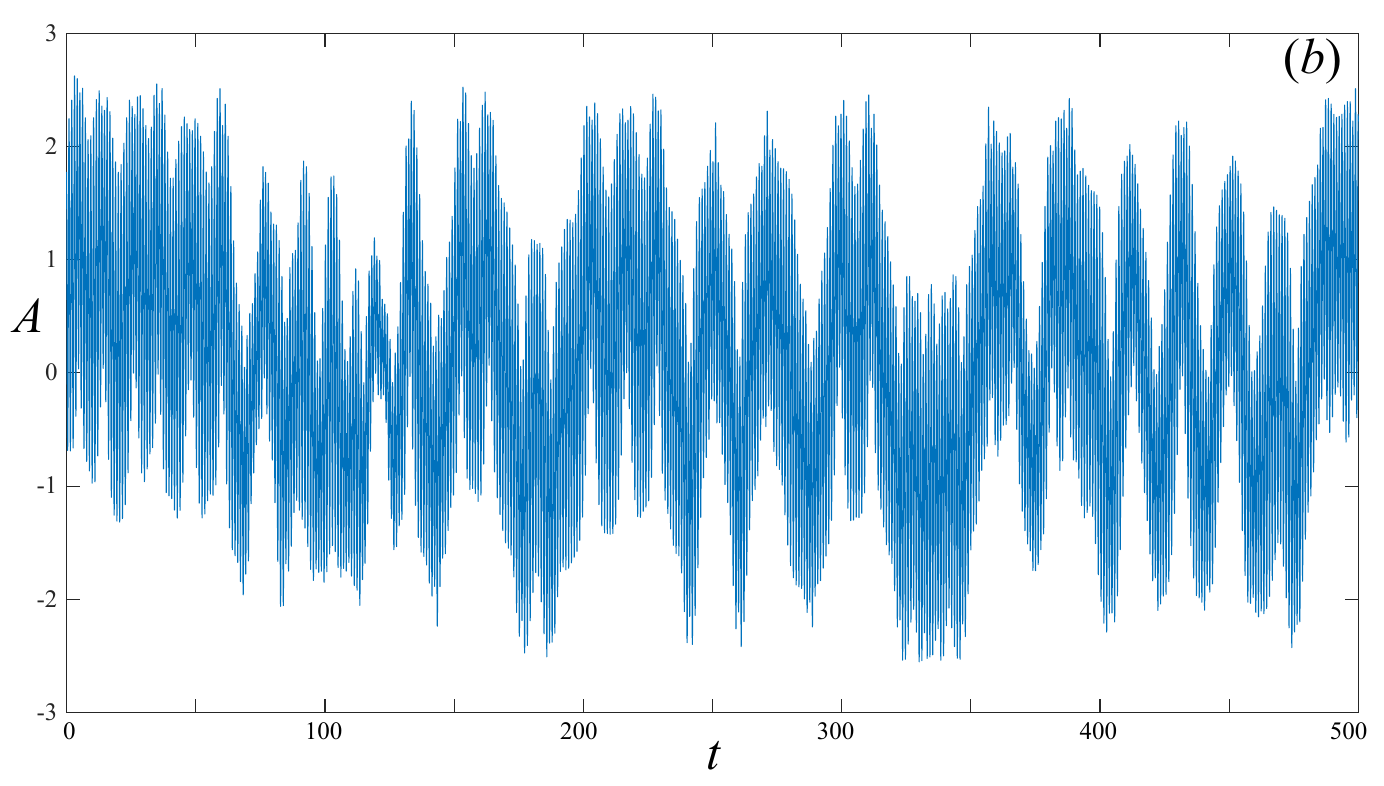}
{For the potential $U(P) = \tfrac12 P^2$ and the initial condition of \Fig{SidesPhase}, 
the perimeter (a) and area (b) as a function of time for $t \in [0,500]$.
For this trajectory segment, $P \in [1.138305,7.985506]$, $A \in [-2.552623,2.623263]$
and the moment of inertia \Eq{InertiaForm} $I \in [0.144625,8.125064]$.} 
{SidesAreaPerim}{5in}

Note that since $E = K + U$, and the kinetic energy is non-negative, the maximal perimeter
for this quadratic potential is
$
	P \le  \sqrt{2E} .
$
For the initial condition in \Fig{SidesPhase}, the energy is $E = 32.445$, implying that $P \le 8.0554$.
However, for this trajectory the angular momentum \Eq{JTotal} is nonzero, so the maximal perimeter is 
determined by
\beq{minEnergy}
	E = \frac{J_T^2}{2I} + U(P) .
\eeq
For a given perimeter and equal masses, the moment of inertia \Eq{InertiaForm} has 
the bound $I \ge P^2/9$, which is reached
when the triangle is equilateral. Indeed, for the trajectory of \Fig{SidesAreaPerim},
$I_{min} = 0.144625 > P_{min}^2/9 = 0.1439709$.
Using this in \Eq{minEnergy} for the quadratic potential, then gives
\beq{Pmax}
		P_{max} = \sqrt{ E + \sqrt{E^2- 9 J_T^2}} = 7.9788.	
\eeq
The maximum realized up to $t = 500$, as seen in \Fig{SidesAreaPerim}(a), is $99.9\%$ of this value.

For a given perimeter the maximum side length occurs when the triangle collapses
to a line, so we must have sides of length less than $P/2$.
The largest distance from the center-of-mass would occur at a double collision, 
where two of the particles are at a distance $P/6$ from the center-of-mass and the third is at the distance $P/3$.
This gives a version of Hill's region: the trajectory must lie in the disk of radius $\tfrac13 P_{max} \simeq 2.6596$
about the center-of-mass; this disk is shown in \Fig{SidesPhase}(a).

The area along the trajectory is shown in \Fig{SidesAreaPerim}(b). For this triangle, $A(0) = 1.77607$ and $A(t)$
initially decreases, crossing zero when the triangle collapses to a line and then reverses orientation. For a given
perimeter the area has an upper bound given by \Eq{Pmin} in \App{Incenter}, corresponding to the equilateral case. 
Thus, since the perimeter is bounded by \Eq{Pmax}, $|A(t)| \le 3.06289$.
The trajectory for this example only reaches $86\%$ of this bound over the time shown in the figure.

We computed the maximal Lyapunov exponent, $\mu$, for this trajectory by integrating the one-jet equations \Eq{oneJet}.
As an initial condition, we chose a random deviation vector that left the center of mass at the origin.
For $T = 2000$, we compute $\mu_T = 0.16905 \pm  0.00008$, where the nominal error is measured by the variation in $\mu_T$
over the interval $T \in [1600,2000]$. As usual it is difficult to give an accurate value for $\mu$  \cite{Sander25b},
but at least this indicates that the trajectory is chaotic.

The variation of the maximal Lyapunov exponent with changing initial velocity is shown in \Fig{LyapunovSides}. For
this one-parameter family of trajectories, $\mu_{2000}$ appears to be positive. 
The three curves in this figure show the value of the exponent with different, randomly chosen,
initial deviation vectors; this gives an indication of the large errors in the computation.

\InsertFig{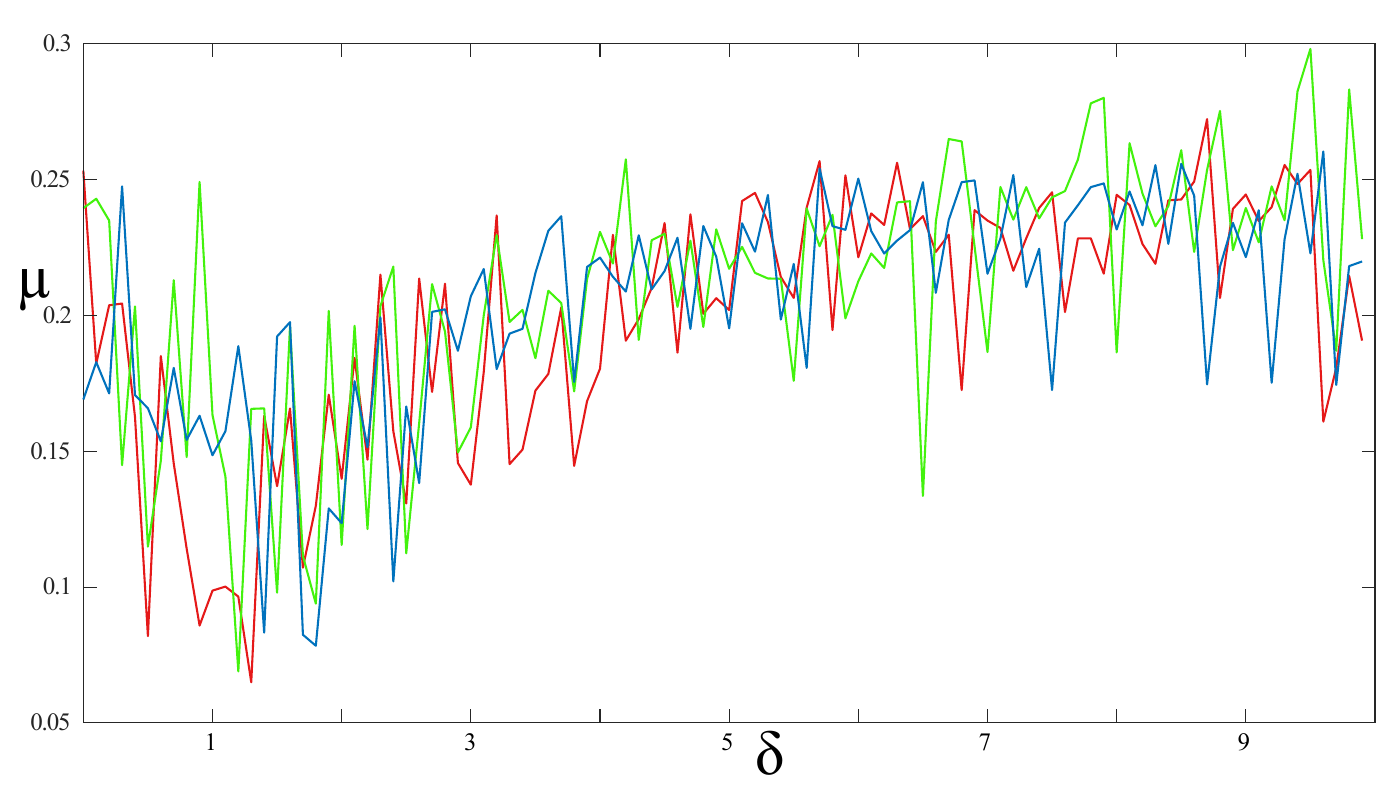}
{Computations of the maximal Lyapunov exponent $\mu_T$ for the potential $U(P) = \tfrac12 P^2$ and
integration time $T = 2000$.
The initial condition is a triangle with sides $(a,b,c) = (2.9,2.0,1.8)$ and initial
velocities $\dot{u} = -\dot{w} = (1+\delta,3)$ and  $\dot{v} = 0$. When $\delta = 0$, this corresponds to the case
shown in \Fig{SidesPhase}.
The three curves differ only in that different randomly chosen deviation vectors were used. Large variations show
the unreliability of the Lyapunov exponent.
}
{LyapunovSides}{5in} 

We now consider a trajectory that is close to the uniformly rotating equilateral triangle $a=b=c$
of \Sec{Special}. Such a trajectory simply rotates when the velocities have magnitude
\[
	 V_{rot} = \sqrt{a U'(P)};
\]
which for the quadratic potential becomes $V_{rot} = \sqrt{3}a$. 
An example with the velocities perturbed is shown in \Fig{EquilateralPhase}. In this case the trajectory 
appears to lie on a three torus.\footnote
{Since we have not eliminated the angular momentum here, we nominally have six degrees of freedom.}
 Indeed the maximal Lyapunov exponent for this case seems to be $0$: $\mu_{200} = 0.013 \pm 0.008$
 and $\mu_{2000} =  0.0029 \pm 0.0008$.
Both the area $A \in [1.477951,5.200333]$ and perimeter $P \in [6.984039,11.026097]$ 
oscillate quasiperiodically (not shown).

\InsertFigTwo{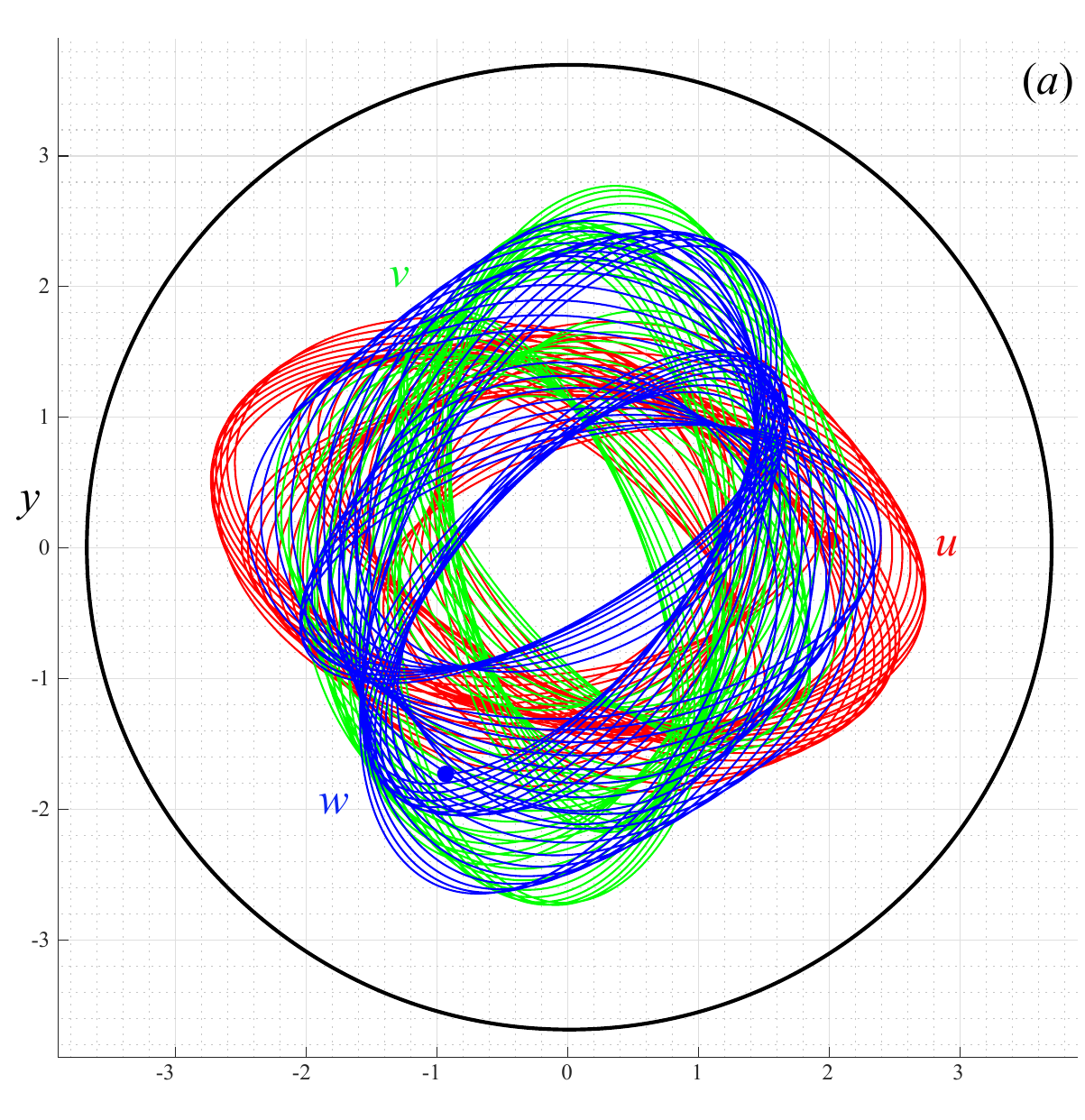}{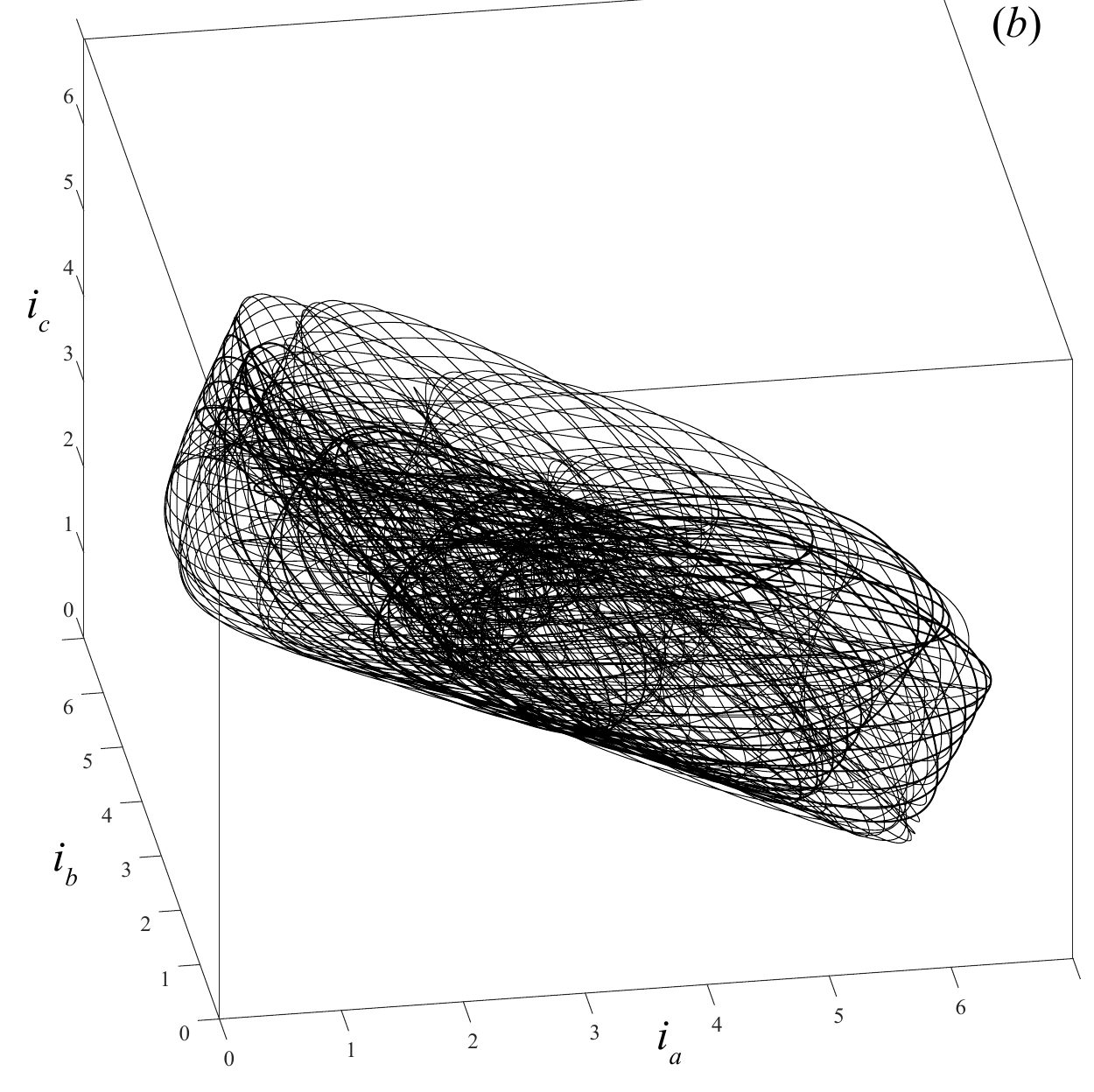}
{A perturbed equilateral triangle.
(a) Positions $(u,v,w)$ in the plane as a function of time, 
up to $t = 100$ for $U(P) = \tfrac12 P^2$. Here the initial condition is an equilateral
triangle with sides $a=b=c= 2\sqrt{3}$.
(b) Trajectory in the positive octant of the inequities, $(i_a,i_b,i_c)$, for $t= 200$..
The initial conditions are
 $u_0 = (2,0)$, 
 $v_0 = (-1, \sqrt{3})$ and 
 $w_0 = (-1, -\sqrt{3})$, with  
$\dot{u}_0 = V_{rot}(0,1) + \delta(\tfrac12,-1)$,
$\dot{v}_0 = \tfrac12 V_{rot}(-\sqrt{3},1)$ and 
$\dot{w}_0 = -\dot{u}_0 - \dot{v}_0$.
The perturbation is $\delta = 3$ and  $V_{rot} = 6$.
For this trajectory, $E = 84.45577$, $J_T = 24.40192$, and $P_{max} = 11.2503$, \Eq{Pmax}.}
{EquilateralPhase}{3in}

The Lyapunov exponent for a range of equilateral initial conditions (those in \Fig{EquilateralPhase} with $\delta \in [0,10]$)
is shown in \Fig{LyapunovScanEquil}. The estimated Lyapunov exponent appears to be essentially zero up to $\delta = 4.2$, 
where there is a sudden onset of chaotic behavior. The short time behavior for several of these regular orbits
is shown in \Fig{EquilateralSet} in inequity space. 
The inner three orbits in the figure appear to lie on invariant tori enclosing the uniformly rotating equilateral case
(the point $i_a = i_b = i_c = 2\sqrt{3}$).
The outermost (gray) trajectory, with $\delta = 4.5$, is chaotic as seen in \Fig{LyapunovScanEquil}.
A longer segment of a trajectory in this family, for $\delta = 3.0$,
was shown in \Fig{EquilateralPhase}.

The  $\delta = 4.5$, chaotic orbit of \Fig{EquilateralSet} appears to have near pair-collisions since it approaches the axes in
inequity space. The first near collision is at $t = 0.498$, where $b = 0.0924$, so 
$u \approx w$; this is followed by a several near collisions of $u$ and $v$, and then a near collision of $v$ and $w$
at $t =  20.212$, where $a = 0.0440$. The evolution of the side lengths for this trajectory is shown in \Fig{EquilPertSides} for $t \in [0,50]$. Of the many other near collisions, the closest, for $t \le  500$, is $a(30.790) = 0.0004372$.
%
%

\InsertFig{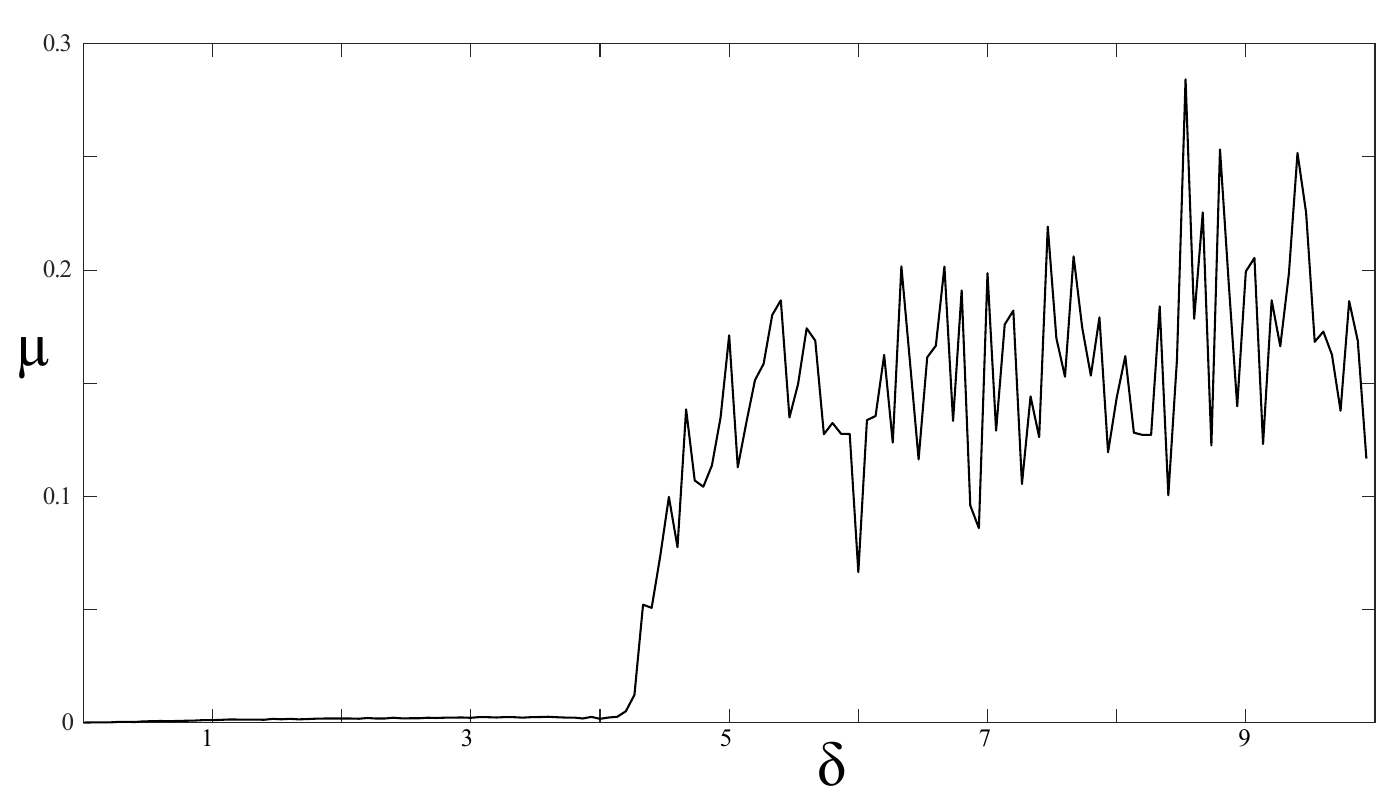}
{ The maximal Lyapunov exponent $\mu_{2000}$ for the equilateral initial condition of \Fig{EquilateralPhase},
with $\dot{u}(0)$ perturbed with $\delta \in [0,10]$. The exponent appears to be zero up to $\delta \approx 4.25$.}
{LyapunovScanEquil}{5in}

\InsertFig{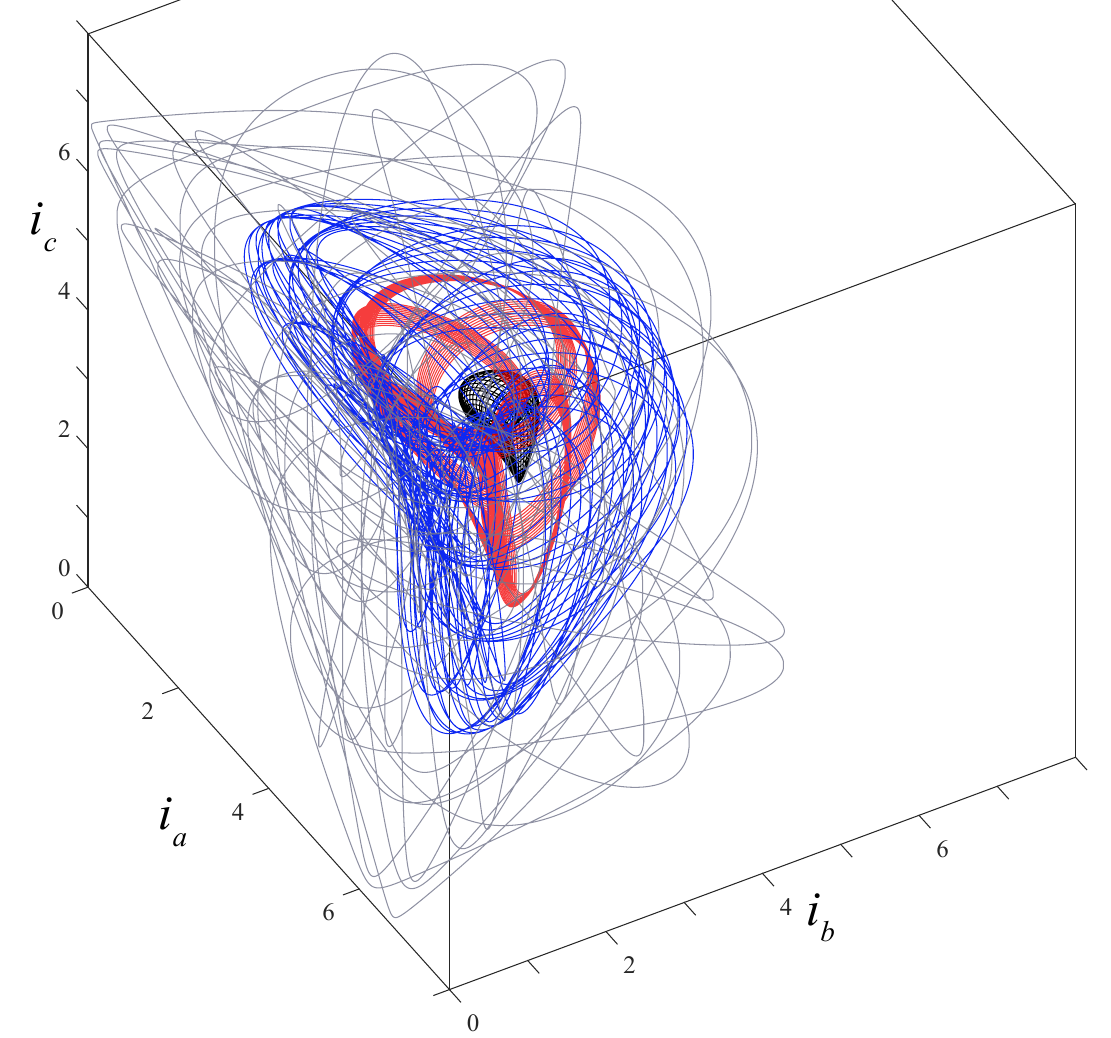}
{Four trajectories in the space of inequities, up to $t = 50$ for the equilateral initial condition of \Fig{EquilateralPhase}, with $\delta = 0.5$ (black), $1.5$ (red), $3.5$ (blue), and $4.5$ (gray).
When $\delta = 0$, the orbit is an equilibrium in this space at $i_a = i_b = i_c = a = 2\sqrt{3}$.
but is a uniformly rotating equilateral triangle in $\bR^2$. The outermost orbit is chaotic
since $\mu \approx 0.176 > 0$.}
{EquilateralSet}{4in}

\InsertFig{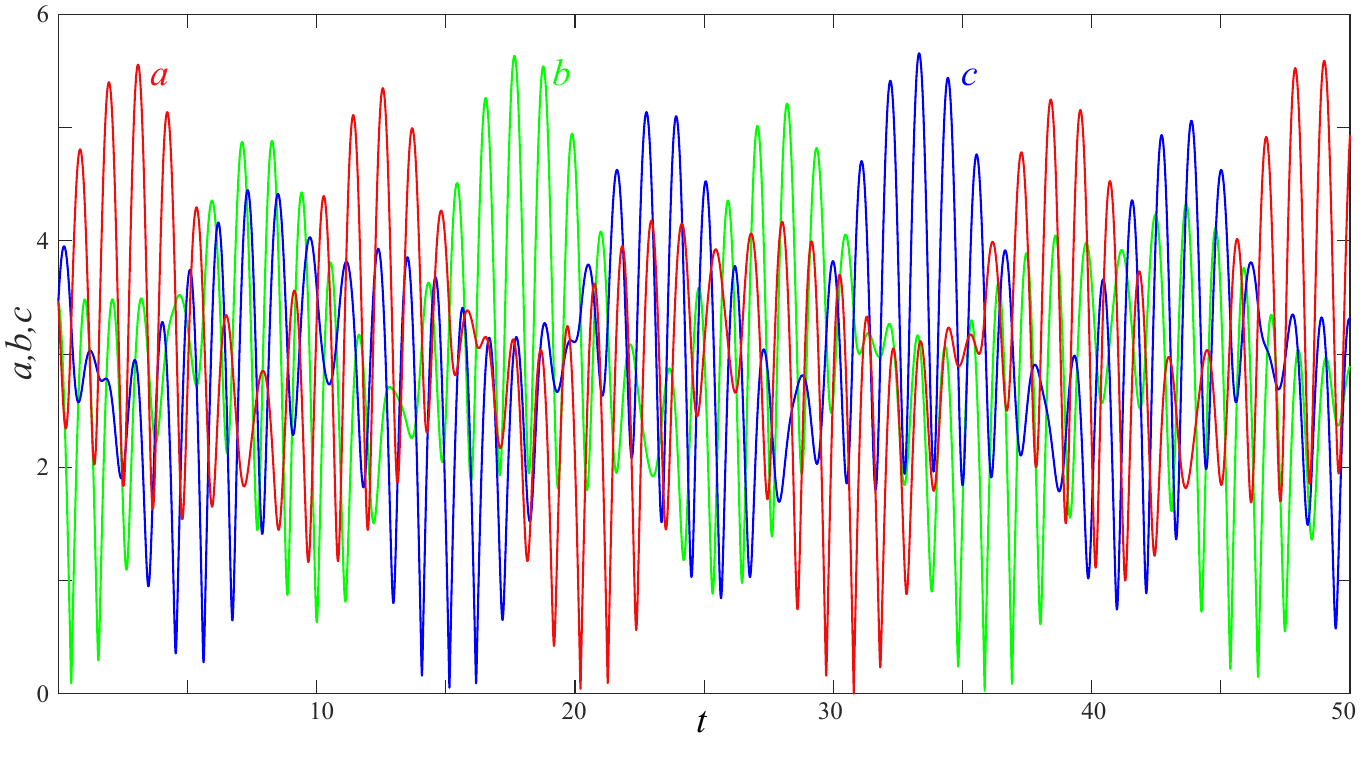}
{The evolution of the triangle sides lengths $a,b,c$ for the initial condition of 
\Fig{EquilateralPhase} with $\delta = 4.5$.
This trajectory was shown as the gray curve in \Fig{EquilateralSet}. Note that there is a sequence of near pair-collisions, when a side length nears zero.}
{EquilPertSides}{4in}

Similar results are obtained for trajectories near the linear case discussed in \Sec{Special}.
Here we consider the set of initial conditions
\bsplit{LinearIC}
	u(0) &= (-2.5, 0), \quad v(0) = (0,0), \quad w(0) = (2.5,0) ,\\
	 \dot{u}(0) &= (1.5, -10.0711), \quad 
	\dot{v}(0) = -\delta (0.5,1.0), \quad 
	\dot{w}(0) = -\dot{u}(0) -\dot{v}(0) ,
\esplit
for $\delta \in (0,10]$. Most orbits in this family appear to have positive Lyapunov exponents, 
with values $0.006 \le \mu_{2000} \le 0.25$.



Finally, we consider an example of a trajectory for the areal case of \Sec{Areal}, using the potential
\beq{ArealExPotential}
	U(\|A\|) =\frac{2}{\|A\|} +  \|A\|^2.
\eeq
Since this is singular at $\|A\| = 0$ orbits cannot collapse to a linear configuration.
For \Fig{ArealSidesPhase}, 
the initial condition is equilateral, but the initial velocities do not have the symmetry \Eq{EquilateralSym}.
During the early evolution, seen in \Fig{ArealSidesPhase}(a), $\|v(t)\|$ (green curve) remains relatively small,
while the other two particles experience larger displacements.
As seen in the inequity plot, \Fig{ArealSidesPhase}(b), 
this orbit repeatedly passes close to the equilateral configuration,
but then undergoes large swings towards either $i_c = 0$ or $i_b = 0$ where the triangle
nears collapse to a linear configuration.
For the areal case, the side lengths can become much larger than the perimetric case:
indeed if the area were to approach zero, the perimeter could be unbounded. 
Of course, this is forbidden by the singularity in the potential \Eq{ArealExPotential};
indeed for this energy, $U(\|A\|) \le E$ restricts the area to the range
\beq{AreaLimits}
	0.3722962812 \le A(t) \le 2.139086602 .
\eeq
The area and perimeter for $t \in [0,500]$ are shown in \Fig{ArealPerimArea}; 
here the area has range $A(t) \in [0.37265, 2.13785]$, within $0.1\%$ of the limits \Eq{AreaLimits}.
The area undergoes much more
rapid oscillations than the perimeter, which has the range $P(t) \in [3.68511, 81.24522]$.
The corresponding moment of inertia (not shown) ranges over $I(t) \in [1.52569,  830.259]$. 
This again obeys the bound $I_{min} \ge P_{min}^2/9 = 1.50889$.

\InsertFigTwo{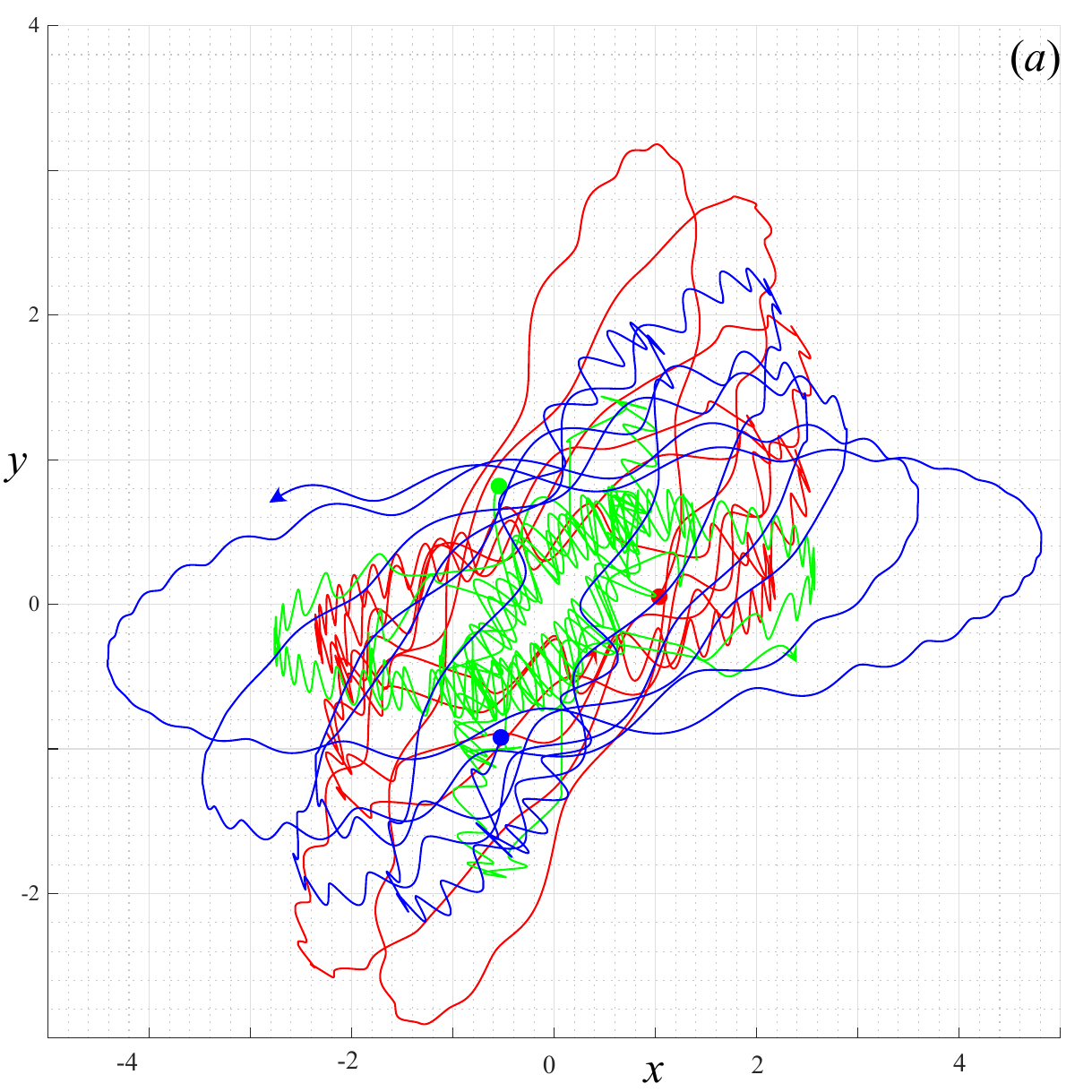}{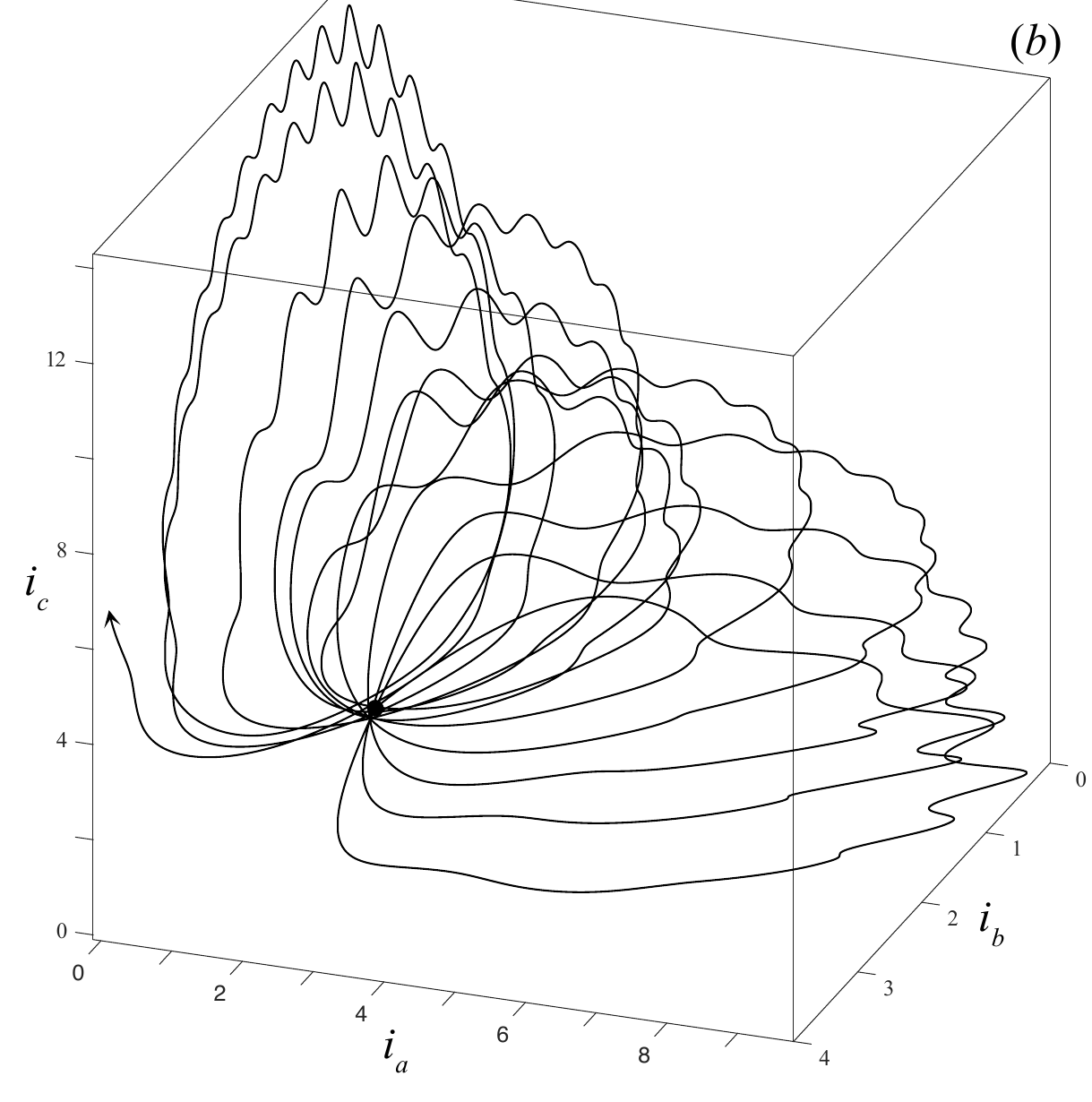}
{(a) Positions of the three particles $(u,v,w)$ in the plane as a function of time, 
up to $t = 100$ for the areal potential \Eq{ArealExPotential}.
(b) Trajectory in the positive octant of the inequities, $(i_a,i_b,i_c)$.
At $t = 0$, the triangle is equilateral with sides $a=b=c= \sqrt{3}$, setting
$u_0  = (1,0))$, 
 $v_0 =\tfrac12 (-1,\sqrt{3})$ and 
 $w_0  =\tfrac12 (-1,-\sqrt{3})$,
The initial velocities are $\dot{u} = -\dot{w} = (1,3)$ and  $\dot{v} = 0$.
For this trajectory, $E =  5.51067$ and $J_T = 1.99322$.}
{ArealSidesPhase}{3in}

\InsertFigTwoVert{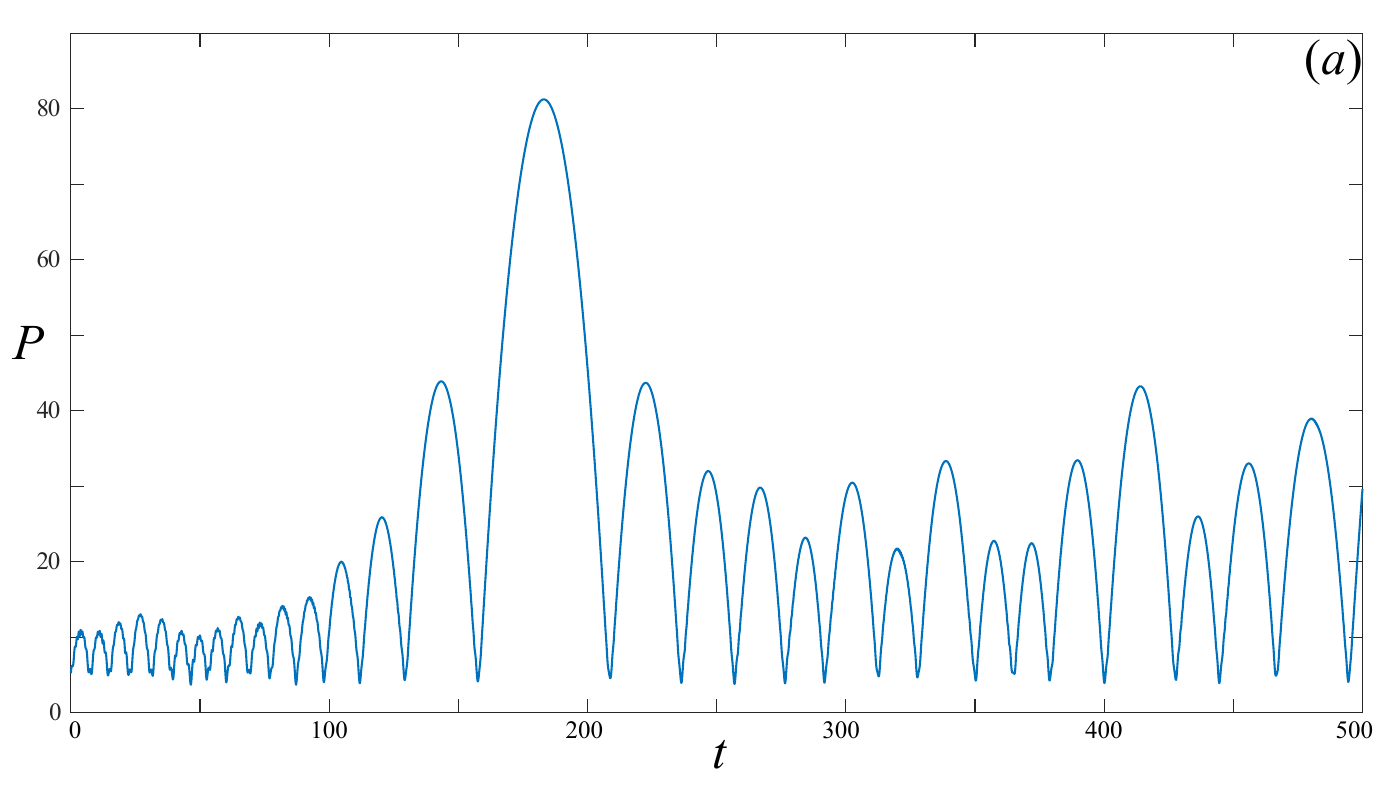}{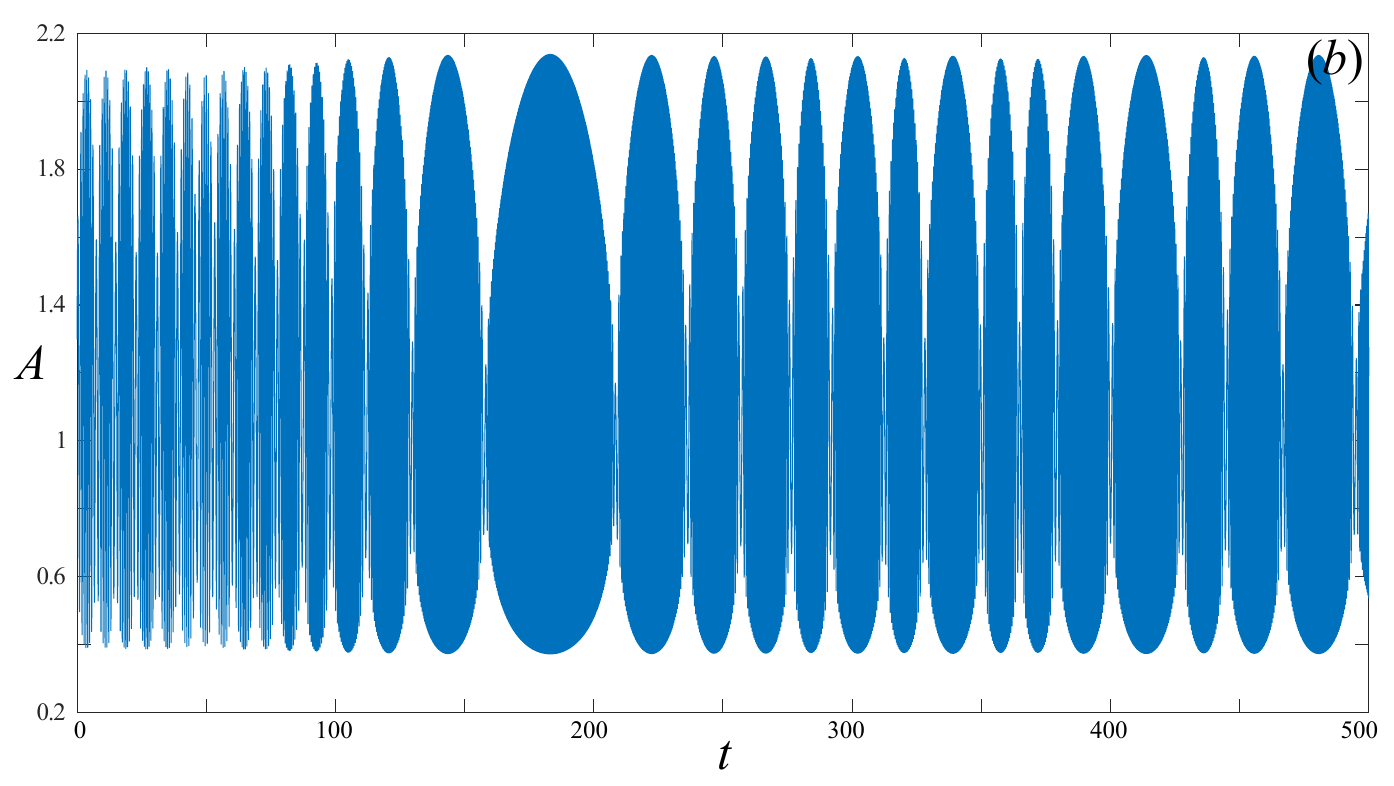}
{For the potential \Eq{ArealExPotential} and the initial condition of \Fig{ArealSidesPhase}, 
the perimeter (a) and area (b) as a function of time for $t \in [0,500]$.} 
{ArealPerimArea}{5in}
\section{Conclusions}\label{Conclusions}
This paper has initiated the study of a rich new class of Hamiltonian systems: particles interacting in triplets with central forces. Geometrically such dynamics is interesting because the forces depend  upon the shape of the triangle.
Analytically, it is unconventional since physical models most often involve pair interactions.
Nevertheless, as we noted in \Sec{Applications}, triplet interactions also arise a number of  physical systems. However, in most of these cases the three-body force is added to a more conventional two-body force.
In our examples, we assumed there were only three-body forces.

In \Sec{ThreeBodies} we  obtained the equations when the potential depends on the perimeter or area of the triangle.
Perimetric potentials also occur in some applications; for example, a potential for colloids was seen in \Eq{Yukawa}.
Potentials that obey the hypothesis of ``centrality", \Hyp{central}, could depend more generally on the side lengths,
but if we ask that they are symmetric, \Hyp{symmetry}, a dependence on the perimeter or area seems natural. We
are not aware of applications in which the potential has the areal form, though such forces have been studied
in the context of gradient dynamics on hypergraphs as we noted in \Sec{Applications}. In \Sec{Simulations}
we obtained numerical solutions for two simple examples, but in other explorations, we found that 
the dynamics has similar behavior for other potentials.
The areal case has a weak singularity if the configuration passes through linearity, where the area is zero;
moreover as in the gravitational case, the triplet force can be singular when pairs or triplets coalesce.
We have leave the study of such singularities to a future paper.

We have found several special solutions: rotating equilateral triangle and linear configurations, and a non-rotating isosceles case. Many trajectories near the equilateral case lie on invariant tori,
but they can become chaotic as the initial conditions vary---a typical phenomenon in Hamiltonian systems
near an elliptic equilibrium. It would be interesting to see if there are other uniformly rotating
solutions that might be analogous to the Lagrange points or central configurations of the gravitational problem \cite{Hampton19}.
For such a study the Jacobi coordinates  and its extensions \cite{Littlejohn98} 
will prove useful. While we obtained this reduction in \Sec{Jacobi}, 
we used a simpler convenient set of reduced coordinates, the triangular inequities, for visualization.
An advantage of the later coordinates is that they respect the symmetry of the triplet system.

There is much to do in the future. For example, it would be interesting to study the 
case when there are more than three particles, as well as to allow for two body forces.
It would also be interesting to investigate in more detail the stability of special solutions and
families of invariant tori that generically occur in Hamiltonian systems due to KAM theory \cite{Meyer92}.

\bigskip
\appendix
\begin{center}
\Large{\textbf{Appendices}}
\end{center}

\section{Triangle Relations}\label{app:Triangles}

Given a triangle with vertices $(u,v,w) \in \bR^9$, we let $(a,b,c) \in \bR^{3+}$ \Eq{Sides}
denote the lengths of each of the opposite sides, recall \Fig{TriangleNotation}.
The altitude of the triangle from the side $c$ to $w$  has length
\beq{Height}
	h_c=  a\sin(\theta_v) =b\sin(\theta_u); 
\eeq
where $\theta_u$ and $\theta_v$ are the interior angles at $u$ and $v$ respectively.
This can be made more explicit by eliminating the sine functions.
 Define the signed distance between $v$ and the base point of the altitude $h_c$
by $x_c = a \cos(\theta_v) = c- b \cos(\theta_u)$, recall \Fig{TriangleNotation}.
Squaring \Eq{Height} then gives 
\beq{xcsolve}
	a^2(1-\cos^2(\theta_v)) = b^2(1-\cos^2(\theta_u)) \quad \Rightarrow \quad
	c^2 + a^2-b^2 = 2 c x_c .
\eeq
With permutations, this gives the cosines of the interior angles:
\bsplit{Cosines}
	\cos(\theta_u) &= \frac{b^2+c^2-a^2}{2bc} , \\
	\cos(\theta_v) &= \frac{c^2+a^2-b^2}{2ca} , \\
	\cos(\theta_w) &= \frac{a^2+b^2-c^2}{2ab} .
\esplit
By the triangle inequalities, the right hand sides of these formulas
are indeed in the interval $[-1,1]$. 


By elementary trigonometry the magnitude of  the area \Eq{Area} is
\[
 	 \|A(u,v,w)\| =\tfrac12  a c \sin(\theta_v) = \tfrac12 c h_c .
 \]
We can solve for the altitude using $h_c^2 = a^2 - x_c^2$ 
and \Eq{xcsolve} to obtain
\[
	h_c =  \frac{1}{2c} \sqrt{(a+b+c)(b+c-a)(c+a-b)(a+b-c)} .
\]
This can be used with \Eq{Height} to give formulas for the sines.
Using this in the area gives Heron's formula
\beq{Heron}
	\|A\| =  \tfrac{1}{4} \sqrt{(a+b+c)(b+c-a)(c+a-b)(a+b-c)} .
\eeq
Note that \Eq{Heron} implies that the area depends only on the lengths
of the sides, and, of course, is permutation symmetric.
%

The momenta of inertia, for the equal mass case is
\[
	I = \|u\|^2 + \|v\|^2 + \|w\|^2
\]
Suppose that the coordinates are chosen so that the center-of-mass is at the origin, $R = \tfrac13 (u+v+w) = 0$, then after some algebra we can see that
\[
	I = \tfrac13 ( a^2 + b^2 + c^2) .
\]
More generally when the masses are not equal, this becomes \Eq{InertiaForm}.
For a given perimeter $P = a+b+c$, this implies that $I \ge P^2/9$, with the bound occurring for the 
equilateral case.
\section{Triangle Inequities}\label{app:Inequities}
For a triangle with sides \Eq{Sides}, the
triangle inequality implies that $|a-b| \le c \le a+b$. We can use this, and its permutations, to define three nonnegative
quantities that we will call the \textit{inequities} of the triangle:
\bsplit{inequities}
	i_a &= b+c-a , \\
	i_b &= c+a-b, \\
	i_c &= a+b-c .
\esplit
In terms of the inequities, the space of all possible triangles is the positive octant
$(i_a,i_b,i_c) \in \bR^{3+}$.
Note that the inequities determine the sides:
\[
	a = \tfrac12 (i_b+i_c), \quad b = \tfrac12 (i_c+i_a), \quad c = \tfrac12 (i_a+i_b) ;
\]
thus they uniquely determine the \textit{shape} of the triangle.
The perimeter \Eq{Perimeter} and area \Eq{Heron} become
\beq{PerimAreaI}
	P = i_a+i_b+i_c , \quad
	\|A\| = \tfrac{1}{4} \sqrt{P i_a i_b i_c} .
\eeq
Surfaces of constant perimeter and area in inequity space are sketched in \Fig{Inequities}.

\InsertFig{Inequities}
{Shape space of triangles in terms of the inequities \Eq{inequities}.
A surface of constant area, $A=3$, is shaded. A partially dashed triangle outlines the plane where $P = 6$;
this plane does not intersect the $A=3$ surface, since \Eq{Pmin} requires $P \gtrsim 7.8964$.
Curves on the $A=3$ surface with perimeters $P = 8, 8.5,$ and $9$ are also shown.
Degenerate triangles are indicated on the coordinate planes.}
{Inequities}{3in}

Each of the planes where one inequity vanishes corresponds to a degenerate triangle:
\begin{align*}
	\{i_a = 0\}& \simeq \{ a = b+c \} \Rightarrow \mbox{ Line with $u$ between $v$ and $w$ } ,\\
	\{i_b = 0\}& \simeq  \{ b = c+a \}  \Rightarrow \mbox{ Line with $v$ between $w$ and $u$ } , \\
	\{i_c = 0\}& \simeq  \{ c = a+b \}  \Rightarrow \mbox{ Line with $w$ between $v$ and $u$ } .
\end{align*}
Examples are shown in \Fig{Inequities}. Thus the axes correspond to pair collisions,
\begin{align*}
	\{i_a = i_b = 0\} & \simeq  \{c = 0\}  \Rightarrow u = v , \\
	\{i_b = i_c = 0\} & \simeq  \{a = 0\}  \Rightarrow v = w ,\\
	\{i_c = i_a = 0\} & \simeq  \{b = 0\}  \Rightarrow w = u .
\end{align*}
Isosceles triangles lie on planes:
\begin{align*}
	\{i_a = i_b\}& \simeq  \{ a = b \} , \\
	\{i_b = i_c\}& \simeq  \{ b = c \} , \\
	\{i_c = i_a\}& \simeq  \{ c = a \} ,
\end{align*}
and equilateral triangles are on the diagonal $i_a=i_b=i_c$. By \Eq{PerimAreaI}, the plane of constant perimeter
intersects the surface of constant area only when 
\beq{Pmin}
	P^2 \ge 12 \sqrt{3} \|A\|, 
\eeq
and the first intersection is on the diagonal where the triangle is equilateral.

\section{Incenter}\label{app:Incenter}
The \textit{incenter}, $c_i$, of a triangle is the center of the inscribed circle defined by the triangle, recall \Fig{PerimeterForces}. It is also the unique point at which bisectors of the three angles meet.
For a triangle with vertices $(u,v,w)$, the bisecting line from a vertex $u$ is 
\[
	l_u(t) = u + t\left( \frac{v-u}{\|v-u\|} + \frac{w-u}{\|w-u\|} \right), \; t \in \bR , 
\]
with cyclic permutations for the two other vertices.
Solving for the intersection point $c_i = l_u(t_1) = l_v(t_2) = l_w(t_3)$ gives
the incenter
\Eq{incenter}, where $P$ is the perimeter \Eq{Perimeter}.

The point $c_i$ is also the center of the inscribed
circle in the triangle: each pair of right triangles formed from the triangle edges and the force line is congruent,
so the three yellow edges in \Fig{PerimeterForces} have equal lengths.
The radius of the incircle is $r_i = \|c_i-u\| \sin(\theta_{u}/2)$,
which, using the trig formulas in \App{Triangles}, becomes
\beq{radius}
	r_i 
	  = 2 \frac{ \|A \|}{P} .
%
\eeq

There are two cases in which the incenter is at the center-of-mass of the triplet. 
If the particles are not collinear, we require that the opposites sides be proportional to the particle masses:
\[
	\|u-v\| = \alpha m_w , \quad
	\|v-w\| = \alpha m_u , \quad
	\|w-u\| = \alpha m_v ,
\]
for then \Eq{incenter} gives
\[
	c_i \to \alpha \frac{m_w w + m_u u + m_v v}{\alpha (m_w + m_u+m_v)} = x_{cm} .
\]
Thus for the equal mass case, this occurs with an equilateral triangle.

If the particles are collinear, then the perimeter is twice maximum interparticle distance. Supposing,
e.g., that $v$ is between $u$ and $w$, then $P = 2\|w-u\| = 2b$. In this case the incenter
is the position of the central particle $c_i =  v$.

\section{Orthocenter}\label{app:Orthocenter}
The orthocenter, $c_0$ of a triangle is the intersection of the altitudes of the triangle $(u,v,w)$,
as shown in \Fig{ArealForces}. To
see that these lines intersect, parallel translate each side to its opposite vertex,
to construct a new triangle $(\mu,\nu,\omega)$, see \Fig{Orthocenter}.
The claim is that $(u,v,w)$ is the medial triangle of $(\mu,\nu,\omega)$: the original vertices are bisectors of the 
sides of the new triangle. Moreover the circumcenter of $(\mu,\nu,\omega)$ is at the position
of the orthocenter of the original triangle.

Solving for the position gives
\[
	c_o 
	   =  v +  \frac{(u-v)\cdot(v-w)}{2 \|A\|} \hat{A} \times (w-u)
\]
or any cyclic permutation.
\InsertFig{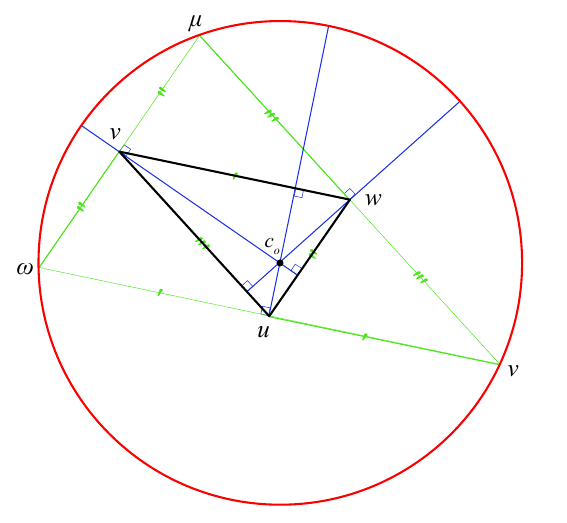}{Construction of the orthocenter of a triangle $(u,v,w)$. The
sides of the triangle $(\mu,\nu,\omega)$ are parallel to each of the sides of the original triangle with
the vertices $(u,v,w)$ as bisectors.
Thus the altitudes of $(u,v,w)$ lie on the radii of the circumcircle of $(\mu,\nu,\omega)$.}
{Orthocenter}{3in}

\section{OneJet}\label{app:OneJet}
To compute the Lyapunov exponent for the perimetric case we use the one-jet
of the vector field \Eq{PerimEq}:
\bsplit{oneJet}
 \delta \dot{p}_u &= U''(P) \left( \frac{v-u}{c} + \frac{w-u}{b} \right) \delta P 
       + U'(P)\left( S^{v-u}_\perp \frac{\delta v -\delta u}{c} +
                      S^{w-u}_\perp \frac{\delta w - \delta u}{b} \right) , \\
 \delta \dot{u} &= \frac{\delta p_u}{m_u} ,
\esplit
where 
\begin{align*}
	\delta P &=  \tfrac{1}{c} (u-v)\cdot (\delta u -\delta v) + \tfrac{1}{b}(u-w)\cdot (\delta u -\delta w) 
	+ \tfrac{1}{a}(v-w)\cdot(\delta v -\delta w) \\
	S_\perp^x &=  I - \frac{x x^T}{\|x\|^2} 
\end{align*}
are the perimeter variation and the matrix projection orthogonal to a vector $x$.
		
\bibliographystyle{alpha}
\bibliography{ThreeBody}
\end{document}